\documentclass[12pt, onecolumn]{IEEEtran}
\usepackage{amsmath}
\usepackage{amssymb}
\usepackage{mathrsfs}
\usepackage{cite}
\usepackage{epsfig}
\usepackage{epsf}
\usepackage{theorem}
\usepackage{graphics}
\usepackage[active]{srcltx}

\renewcommand{\baselinestretch}{1.62}

\begin{document}

\title{Green Modulation in Proactive Wireless Sensor Networks \\
\thanks{The material in this paper was submitted in part to ICASSP'2010 conference, September 2009 \cite{Jamshid_ICASSP2010}.}}

\author{
\small  Jamshid Abouei, \emph{Member, IEEE}, Konstantinos N. Plataniotis, \emph{Senior Member, IEEE} and Subbarayan Pasupathy, \emph{Fellow, IEEE}\\
\small The Edward S. Rogers Sr. Department of ECE, \\
University of Toronto, Toronto, ON, M5S 3G4, Canada, \\
Emails:~ \{abouei, kostas and pas\}@comm.utoronto.ca.}

\maketitle

\begin{abstract}

Due to unique characteristics of sensor nodes, choosing
energy-efficient modulation scheme with low-complexity
implementation (refereed to as \emph{green modulation}) is a
critical factor in the physical layer of Wireless Sensor Networks
(WSNs). This paper presents (to the best of our knowledge) the first
in-depth analysis of energy efficiency of various modulation schemes
using realistic models in IEEE 802.15.4 standard and present
state-of-the art technology, to find the best scheme in a proactive
WSN over Rayleigh and Rician flat-fading channel models with
path-loss. For this purpose, we describe the system model according
to a pre-determined time-based process in practical sensor nodes.
The present analysis also includes the effect of bandwidth and
active mode duration on energy efficiency of popular modulation
designs in the pass-band and Ultra-WideBand (UWB) categories.
Experimental results show that among various pass-band and UWB
modulation schemes, Non-Coherent M-ary Frequency Shift Keying
(NC-MFSK) with small order of $M$ and On-Off Keying (OOK) have
significant energy saving compared to other schemes for short range
scenarios, and could be considered as realistic candidates in WSNs.
In addition, NC-MFSK and OOK have the advantage of less complexity
and cost in implementation than the other schemes.
\end{abstract}

\begin{center}
\end{center}

\section{Introduction}
Wireless Sensor Networks (WSNs) represent a new generation of
distributed systems to support a broad range of applications;
including monitoring and control, health care, tracking
environmental pollution levels, and traffic coordination. A WSN
consists of a large number of microsensors which are typically
powered by small energy-constrained batteries. In many application
scenarios, these batteries can not be replaced or recharged, making
the sensor useless once battery life is over. Thus, minimizing the
total energy consumption (i.e., circuits and signal transmission
energies) is a critical factor in designing a WSN. Due to the unique
characteristics and application requirements of sensor networks,
deploying energy-efficient design guidelines in traditional wireless
networks are not suited to the WSNs. In fact, since the output
transmission energy is dependent to the distance between transmitter
and receiver, long distance wireless communication systems use
efficient protocols (emphasizing on modulation schemes) to minimize
the output transmission energy. In these systems, the energy
consumed by circuits is ignored. However, typical WSNs assume that
sensor nodes are densely deployed which means distance between
sensor nodes is normally short compared to the traditional wireless
systems. Thus, the circuit energy consumption in a WSN is comparable
to the output transmission energy. For energy-constrained WSNs, on
the other hand, the data rates are usually low. Thus, using
complicated signal processing techniques (e.g., high order M-ary
modulation and complex detection) are not desirable for low-power
WSNs.

Several energy-efficient approaches have been investigated for
different layers of a WSN \cite{YeINFOCOM2002, KwonITWC1206,
Cui_GoldsmithITWC0905, TangITWC0407, Qu_ITSP0908}. Central to the
study of energy-efficient techniques in physical layer of a WSN is
modulation. Since, achieving all requirements (e.g., minimum energy
consumption, maximum bandwidth efficiency, high system performance
and low complexity) is a complex task in a WSN, and due to the power
limitation in sensor nodes, the choice of a proper modulation scheme
is a main challenge in designing a WSN. Taking this into account, an
energy-efficient modulation scheme should be simple enough to be
implemented by state-of-the-art low-power technology, but still
robust enough to provide the desired service. In addition, since
sensor nodes frequently switch from sleep mode to active mode,
modulation circuits should have fast start-up times. We refer to
these low-energy consumption schemes as \emph{green modulations}. In
recent years, several energy-efficient modulation schemes have been
studied in WSNs \cite{Cui_GoldsmithITWC0905, TangITWC0407,
Qu_ITSP0908, Wang_ISLPED2001, Qu_ICASSP2007, GursoyISIT2008}.
Broadly, they can be divided into pass-band and Ultra-WideBand (UWB)
modulation schemes. Pass-band modulation schemes such as M-ary
Frequency Shift Keying (MFSK), M-ary Quadrature Amplitude Modulation
(MQAM) and M-ary Phase Shift Keying (MPSK) use sinusoidal carrier
signal for modulation. The complexity of receiver circuitry in
traditional M-ary modulation schemes (e.g., MQAM and MPSK) makes
their implementation on WSNs rather costly despite their great
performance, e.g., Bit-Error-Rate (BER). In addition, they require
Digital-to-Analog Converters (DACs) and mixers which are the most
power intensive components at the receiver \cite{EnzRFIT2005}.
%More precisely, traditional M-ary modulation schemes are not necessarily
%energy-efficient for dense WSNs, since the extra energy consumed by
%the circuitry would surpass the energy saved by the signal
%transmission for a given BER.
%MFSK scheme, on the other hand, has
%proven to be the most appropriate implementation scheme in reducing
%circuit cost and complexity in a WSN \cite{Wang_ISLPED2001} ,
%\cite{EnzRFIT2005}.
Ultra-wideband known as digital pulse wireless is a very short-range
wireless technology (typically for 10 meters distance or less
\cite{Karl_Book2005}) for transmitting data over a wide spectrum.
UWB modulation schemes such as Pulse Position Modulation (PPM) and
On-Off Keying (OOK) require no sinusoidal carrier signal for
modulation. The main advantages of UWB modulation schemes are their
immunity to multipath, very low transmission power and simple
transceiver circuity.

Several energy-efficient modulation schemes have been investigated
in physical layer of WSNs \cite{Wang_ISLPED2001, TangITWC0407,
LiangPACRIM2007, Qu_ICASSP2007, Qu_ITSP0908, Cui_GoldsmithITWC0905,
Karvonen2004, Shen2008, GarzasEURASIP2007}. Tang \emph{et al.}
\cite{TangITWC0407} compare the battery power efficiency of PPM and
FSK schemes in a WSN without considering the effect of bandwidth and
active mode period. Under the assumption of the non-linear battery
model, reference \cite{TangITWC0407} shows that FSK is more
power-efficient than PPM in sparse WSNs, while PPM may outperform
FSK in dense WSNs. References \cite{Qu_ITSP0908} and
\cite{Qu_ICASSP2007} compare the battery power efficiency of PPM and
OOK based on the exact BER and the cutoff rate for WSNs with
path-loss Additive White Gaussian Noise (AWGN) channels. Reference
\cite{GarzasEURASIP2007} investigates the energy efficiency of a
centralized WSN with adaptive MQAM scheme. Most of the pioneering
work on energy-efficient modulation (e.g. \cite{TangITWC0407,
Qu_ITSP0908, Wang_ISLPED2001, Qu_ICASSP2007}) has only focused on
minimizing the average energy consumption of transmitting one bit
without considering the effect of bandwidth and transmission time
duration. In a practical WSN however, it is shown that minimizing
the total energy consumption depends strongly on active mode
duration and the channel bandwidth. Reference
\cite{Cui_GoldsmithITWC0905} addresses this problem only for MFSK
and MQAM for AWGN channel model with path-loss, and shows that MQAM
is more energy-efficient than MFSK for short-range applications. In
addition, some literature (e.g., \cite{TangITWC0407}) have compared
the energy consumption of pass-band modulation schemes with that of
UWB ones without considering the point that the channel bandwidth of
UWB schemes is much wider than that of pass-band ones. Thus, due to
the dependency of total energy consumption on bandwidth, this kind
of comparison would not meaningful. Furthermore deploying adaptive
modulation schemes \cite{GarzasEURASIP2007} are not practically
feasible in WSNs, as they require some additional system complexity
as well as channel state information fed back from sink node to
sensor node.

In this paper, we analyze in-depth the energy efficiency of various
modulation schemes considering the effect of channel bandwidth and
active mode duration to find the green modulation in a proactive
wireless sensor network. For this purpose, we describe the system
model according to a pre-determined time-based process in practical
sensor nodes. Also, new analysis results for comparative evaluation
of popular modulation designs in the pass-band and UWB categories
are introduced according to realistic parameters in IEEE 802.15.4
standard \cite{IEEE_802_15_4_2006}. We start the energy efficiency
analysis based on Rayleigh flat-fading channel model with path-loss
which is a feasible model in static WSNs \cite{TangITWC0407,
Qu_ITSP0908}. Then, we evaluate numerically the energy efficiency of
pass-band modulation schemes operating over the more general Rician
model which includes a strong direct Line-Of Sight (LOS) path.
Experimental results show that among various pass-band modulation
schemes, non-coherent MFSK is a realistic option in short to
moderate range WSNs, since it has the advantage of less complexity
and cost in implementation than MQAM and Offset Quadrature Phase
Shift Keying (OQPSK) used in Zigbee/IEEE 802.15.4 protocols, and has
less total energy consumption. In addition, since for typical
energy-constrained WSNs, data transmission rates are usually low,
using small order M-ary FSK schemes are desirable. Furthermore,
simulation results show that OOK has less total energy consumption
in very short WSN applications, along with the advantage of less
complexity and cost in implementation than M-PPM.

The rest of the paper is organized as follows. In Section
\ref{model_Ch2}, the proactive system model and assumptions are
described. A comprehensive analysis of the energy efficiency for
various pass-band and UWB modulation schemes is presented in
Sections \ref{Analysis_Ch 3} and \ref{Analysis_Ch 4}. Section
\ref{simulation_Ch5} provides some numerical evaluations using
realistic models to confirm our analysis. Also, some design
guidelines for green modulation in practical WSN applications are
presented. Finally in Section \ref{conclusion_Ch6}, an overview of
the results and conclusions is presented.

\section{System Model and Assumptions}\label{model_Ch2}
In this work, we consider a \emph{proactive} wireless sensor system,
in which a sensor node samples continuously the environment and
transmits the equal amount of data per time unit to a designated
sink node. This proactive system is the case of many environmental
applications such as sensing temperature, solar radiation and level
of contamination \cite{Cordeiro_Book2006}. The sensor and sink nodes
synchronize with each other and work in a real time-based process as
depicted in Fig \ref{fig: Time-Basis}. During \emph{active mode}
duration $T_{ac}$, the analog signal sensed by the sensor is first
digitized by an Analog-to-Digital Converter (ADC), and an $N$-bit
message sequence $(a_1,a_2,...,a_N)$ is generated, where $N$ is
assumed to be fixed, and $a_i \in \{0,1 \}$, $i=1,2,...,N$. The
above stream is modulated using a pre-determined modulation
scheme\footnote{Because the main goal of this work is to find green
modulation scheme (i.e., energy-efficient scheme with low complexity
in implementation), and noting that the source/channel coding blocks
increase the complexity of the system, in particular codes with
iterative decoding process \cite[pp. 158-160]{Karl_Book2005}, the
source/channel coding blocks are not considered.} and then
transmitted to the sink node. Finally, sensor node back to
\emph{sleep mode}, and all the circuits of the transceiver is
shutdown in sleep mode duration $T_{sl}$ for energy saving. We
denote $T_{tr}$ as the \emph{transient mode} duration consisting of
the switching time from sleep mode to active mode (i.e., $T_{sl
\rightarrow ac}$) plus the switching time from active mode to sleep
mode (i.e., $T_{ac \rightarrow sl}$), where $T_{ac \rightarrow sl}$
is assumed to be negligible. When a sensor switches from sleep mode
to active mode to send data, a significant amount of power is
consumed for starting up the transmitter, while the power
consumption during $T_{ac \rightarrow sl}$ is negligible. Thus,
choosing an efficient-modulation scheme with fast start-up circuits
is desirable in designing WSNs. Under the above considerations, the
sensor node has $N$ bits to transmit during $0 \leq T_{ac} \leq
T_N-T_{tr}$, where $T_N \triangleq T_{tr}+T_{ac}+T_{sl}$ is assumed
to be fixed for each modulation scheme, and $T_{tr} \approx T_{sl
\rightarrow ac}$. Note that $T_{ac}$ is a critical factor in
choosing an efficient modulation scheme, as it directly affects the
total energy consumption as we will see later.

Since sensor nodes in a typical WSN are densely deployed, the
circuits power consumption $\mathcal{P}_c \triangleq
\mathcal{P}_{ct}+\mathcal{P}_{cr}$ is comparable to the output
transmit power consumption denoted by $\mathcal{P}_t$, where
$\mathcal{P}_{ct}$ and $\mathcal{P}_{cr}$ represent the circuit
power consumptions of sensor and sink nodes, respectively. Taking
these into account, the total energy consumption in the active mode
period, denoted by $\mathcal{E}_{ac}$, is given by
$\mathcal{E}_{ac}=(\mathcal{P}_c+\mathcal{P}_t)T_{ac}$, where
$T_{ac}$ is a function of $N$ and the channel bandwidth as we will
show in Section \ref{Analysis_Ch 3}. Also, the energy consumption in
sleep mode duration, denoted by $\mathcal{E}_{sl}$, is given by
$\mathcal{E}_{sl}=\mathcal{P}_{sl}T_{sl}$, where $\mathcal{P}_{sl}$
is the corresponding power consumption. It is worth mentioning that
during sleep mode period, the \emph{leakage current} coming from
CMOS circuits is a dominant factor in $\mathcal{P}_{sl}$. Clearly,
higher sleep mode duration increases energy consumption
$\mathcal{E}_{sl}$ due to increasing leakage current as well as
$T_{sl}$. Present state-of-the art technology aims to keep a low
sleep mode leakage current no longer than the battery leakage
current, which results in $\mathcal{P}_{sl}$ much smaller than the
power consumption in active mode \cite{Mingoo2007}. For this reason,
we assume that $\mathcal{P}_{sl}\approx 0$. As a result, the
\emph{energy efficiency}, referred to as the performance metric of
the proposed WSN, can be defined as the total energy consumption in
each period $T_N$ correspond to $N$-bit message as follows:
\begin{equation}\label{total_energy1}
\mathcal{E}_N \approx
(\mathcal{P}_c+\mathcal{P}_t)T_{ac}+\mathcal{P}_{tr}T_{tr},
\end{equation}
where $\mathcal{P}_{tr}$ is the circuit power consumption during
transient mode period. We use (\ref{total_energy1}) to investigate
 and compare the energy efficiency of various modulation schemes.

\textbf{Channel Model:} The choice of low transmission power in WSNs
represents several consequences for channel modeling. It is shown by
Friis \cite{Friis1946} that a low transmission power implies a small
range. On the other hand, for short-range transmission scenarios,
the root mean square (rms) delay spread is in the range of
nanoseconds \cite{Karl_Book2005} (and picoseconds for UWB
applications \cite{Tanchotikul2006}) which is small compared to
symbol durations for modulation schemes. For instance, the channel
bandwidth and the correspond symbol duration considered in IEEE
802.15.4 standard are $B=62.5$ KHz and $T_s=16~\mu$s, respectively
\cite[p. 49]{IEEE_802_15_4_2006}, while the rms delay spread in
indoor environments are in the range of 70-150 ns
\cite{Barclay2003}. Thus, it is reasonable to expect a flat-fading
channel model for WSNs. Under the above considerations, the channel
model between the sensor and sink nodes is assumed to be Rayleigh
flat-fading with path-loss, which is a feasible model in static WSNs
\cite{TangITWC0407, Qu_ITSP0908}. We denote the fading channel
coefficient correspond to a transmitted symbol $i$ as $h_{i}$, where
the amplitude $\big\vert h_{i} \big\vert$ is Rayleigh distributed
with probability density function (pdf) given according to $f_{\vert
h_{i} \vert}(r)=\frac{2r}{\Omega}e^{-\frac{r^2}{\Omega}},~r \geq 0$,
where $\Omega \triangleq \mathbb{E}\left[\vert h_{i} \vert^2\right]$
(pp. 767-768 of \cite{Proakis2001}). This results $\vert h_{i}
\vert^2$ being \emph{chi-square} distributed with 2 degrees of
freedom, where $f_{\vert h_{i}
\vert^2}(r)=\frac{1}{\Omega}e^{\frac{-r}{\Omega}}$.

To model the path-loss of a link in a distance $d$, let denote
$\mathcal{P}_t$ and $\mathcal{P}_r$ as the transmitted and the
received signal powers, respectively. For a $\eta^{th}$-power
path-loss channel, the channel gain factor is given by
$\mathcal{L}_d \triangleq
\frac{\mathcal{P}_t}{\mathcal{P}_r}=M_ld^\eta \mathcal{L}_1$, where
$M_l$ is the gain margin which accounts for the effects of hardware
process variations, background noise and $\mathcal{L}_1 \triangleq
\frac{(4 \pi)^2}{\mathcal{G}_t \mathcal{G}_r \lambda^2}$ is the gain
factor at $d=1$ meter which is specified by the transmitter and
receiver antenna gains $\mathcal{G}_t$ and $\mathcal{G}_r$, and
wavelength $\lambda$ (e.g., \cite{Cui_GoldsmithITWC0905},
\cite{Qu_ITSP0908} and \cite{MinISPED2002}). As a result, when both
fading and path-loss are considered, the instantaneous channel
coefficient becomes $G_{i} \triangleq
\frac{h_{i}}{\sqrt{\mathcal{L}_d}}$. Denoting $x_i(t)$ as the
transmitted signal with energy $\mathcal{E}_{t}$, the received
signal at sink node is given by $y_{i}(t)=G_{i}x_{i}(t)+n_{i}(t)$,
where $n_{i}(t)$ is AWGN with two-sided power spectral density given
by $\frac{N_{0}}{2}$. Under the above considerations, the
instantaneous Signal-to-Noise Ratio (SNR), denoted by $\gamma_{i}$,
correspond to an arbitrarily symbol $i$ can be computed as
$\gamma_{i}=\frac{\vert G_{i}\vert^2 \mathcal{E}_t}{N_0}$. Under the
assumption of Rayleigh fading channel model, $\gamma_{i}$ is
chi-square distributed with 2 degrees of freedom, with pdf
$f_{\gamma}(\gamma_{i})=\frac{1}{\bar{\gamma}}\exp\left(
-\frac{\gamma_{i}}{\bar{\gamma}}\right)$, where $\bar{\gamma}
\triangleq \mathbb{E}[\vert
G_{i}\vert^2]\frac{\mathcal{E}_t}{N_0}=\frac{\Omega}{\mathcal{L}_d}\frac{\mathcal{E}_t}{N_0}$
denotes the average received SNR.

\section{Energy Consumption Analysis of Pass-band Modulation Schemes}\label{Analysis_Ch 3}
Pass-band modulation schemes such as MFSK, MQAM and OQPSK use
sinusoidal carrier signal for modulation. In the following, we
investigate three popular carrier-based modulation schemes from
energy and bandwidth efficiency points of view over Rayleigh fading
channel model with path-loss\footnote{ In the sequel and for
simplicity of notation, we use the superscripts `FS', `QA' and `OQ'
for MFSK, MQAM and OQPSK, respectively.}.

\textbf{M-ary FSK:} Let assume the bits stream $(a_1,...,a_N)$ is
considered as modulating signals in an $M$-ary FSK modulation
scheme, where $M$ orthogonal carriers can be mapped into $b
\triangleq \log_{2}M$ bits. The main advantage of this M-ary
orthogonal signaling is that received signals do not interfere with
each other in the process of detection at the receiver. An MFSK
modulator benefits from the advantage of using the Direct Digital
Modulation (DDM) approach, i.e., it does not need mixer and DAC
which are used for MQAM and MPSK (see Fig. \ref{fig: FSKmodulator}).
In fact, MFSK modulation is usually implemented digitally inside the
frequency synthesizer. This property makes MFSK has faster start-up
time than other pass-band schemes \cite{EnzRFIT2005}. The output of
the frequency synthesizer can be frequency modulated and controlled
simply by $b$ bits in the input of a ``digital control" block. The
modulated signal is then filtered again, amplified by the Power
Amplifier (PA), and finally transmitted to the wireless channel.

Denoting $\mathcal{E}^{FS}_t$ as the MFSK transmit energy per symbol
with symbol duration $T^{FS}_s$, the MFSK transmitted signal is
given by
$x^{FS}_{i}(t)=\sqrt{\frac{2\mathcal{E}^{FS}_t}{T^{FS}_s}}\cos(2\pi(f_0+i.\Delta
f)t),~i=0,1,...M-1$, where $f_0$ is the first carrier frequency in
the MFSK modulator and $\Delta f$ is the minimum carrier separation
which is equal to $\frac{1}{2T^{FS}_s}$ for coherent FSK and
$\frac{1}{T^{FS}_s}$ for non-coherent FSK \cite[p. 114]{Xiong_2006}.
Thus, the channel bandwidth $B$ is obtained as $B \approx
M\times\Delta f $, where $B$ is assumed to be fixed for all
pass-band modulation schemes. Denoting $B^{FS}_{eff}$ as the
\emph{bandwidth efficiency} of MFSK (in units of bits/s/Hz) defined
as the ratio of data rate $R^{FS}=\frac{b}{T^{FS}_s}$ (bits/sec) to
the channel bandwidth, we have $B^{FS}_{eff} \triangleq
\frac{R^{FS}}{B} = \frac{\zeta \log_2 M}{M}$, where $\zeta=2$ for
coherent and $\zeta=1$ for non-coherent FSK. It is observed that
increasing constellation size $M$ leads to decrease in the bandwidth
efficiency in MFSK. However, the effect of increasing $M$ on the
energy efficiency should be considered as well. To address this
problem, we first derive the relationship between $M$ and the active
mode duration $T^{FS}_{ac}$. Noting that during each symbol period
$T^{FS}_{s}$, we have $b$ bits, it is concluded that
\begin{equation}\label{active1}
T^{FS}_{ac}=\dfrac{N}{b}T^{FS}_{s}=\dfrac{MN}{\zeta B\log_2 M}.
\end{equation}
Recalling that $B$ and $N$ are fixed, increasing $M$ results in
increasing in $T^{FS}_{ac}$. However, as illustrated in Fig.
\ref{fig: Time-Basis}, the maximum value for $T^{FS}_{ac}$ is
bounded by $T_N-T^{FS}_{tr}$. Thus, the maximum constellation size
$M$, denoted by $M_{max}\triangleq 2^{b_{max}}$, for MFSK is
calculated by the non-linear equation $ \frac{2^{b_{max}}}{\zeta
b_{max}}=\frac{B}{N}(T_{N}-T^{FS}_{tr})$.

At the receiver side, the received MFSK signal can be detected
coherently to provide optimum performance. However, the MFSK
coherent detection requires the receiver to obtain a precise
frequency and carrier phase reference for each of the transmitted
orthogonal carriers. For large $M$, this would increase the
complexity of detector which makes a coherent MFSK receiver very
difficult to implement. Due to the above considerations, most
practical MFSK receivers use non-coherent detectors. The optimum
Non-Coherent MFSK (NC-MFSK) consists of a bank of $M$ matched
filters, each followed by an envelop (or square-law) detector
\cite{ZiemerBook1985}. At the sampling times $t=\ell T^{FS}_s$, a
maximum-likelihood makes decision based on the largest filter
output. It is worth mentioning that using an envelope detector for
demodulation avoids the use of power-intensive active analogue
components such as mixer in coherent detectors.

Now, we are ready to derive the total energy consumption of a
NC-MFSK\footnote{For the purpose of comparison, the energy
efficiency of a \emph{coherent} MFSK is fully analyzed in Appendix
\ref{append01}.}. We first derive $\mathcal{E}^{FS}_{t}$, the
transmit energy per symbol, in terms of a given average Symbol Error
Rate (SER) denoted by $P_{s}$. The average SER of a NC-MFSK is given
by
$P_{s}=\int_{0}^{\infty}P_{s}(\gamma_{i})f_{\gamma}(\gamma_{i})d\gamma_{i}$,
where $P_{s}(\gamma_{i})$ namely the SER conditioned upon
$\gamma_{i}$, is obtained as
$P_{s}(\gamma_{i})=\int_{0}^{\infty}uI_0(\sqrt{2\gamma_{i}u})\left[1-\left(
1-e^{-\frac{u^2}{2}}\right)^{M-1}
\right]e^{-\frac{u^2+2\gamma_{i}}{2} }du$, where $I_0(x)$ is the
zeroth order modified Bessel function \cite{KhalonaITC0496,
TangITWC0407}. It is shown in \cite[Lemma 2]{TangITWC0407} that the
above $P_s$ is upper bounded by $P_{s} \leq
1-\left(1-\frac{1}{2+\bar{\gamma}^{FS}}\right)^{M-1}$, where
$\bar{\gamma}^{FS}
=\frac{\Omega}{\mathcal{L}_d}\frac{\mathcal{E}^{FS}_t}{N_0}$. As a
result, the transmit energy consumption per each symbol is obtained
as $\mathcal{E}^{FS}_t \triangleq \mathcal{P}^{FS}_t T^{FS}_s \leq
\left[\left( 1-(1-P_s)^{\frac{1}{M-1}}\right)^{-1}-2
\right]\frac{\mathcal{L}_d N_0}{\Omega}$. Since, it aims to obtain
the maximum energy consumption, we approximate the above upper bound
as an equality. Using (\ref{active1}), the output energy consumption
of transmitting $N$ bits during $T^{FS}_{ac}$ is computed as
\begin{equation}\label{energy_trans1}
\mathcal{P}^{FS}_t T^{FS}_{ac} =
\dfrac{T^{FS}_{ac}}{T^{FS}_s}\mathcal{E}^{FS}_t \approx \left[\left(
1-(1-P_s)^{\frac{1}{M-1}}\right)^{-1}-2 \right]\dfrac{\mathcal{L}_d
N_0}{\Omega} \dfrac{N}{\log_2 M},
\end{equation}
which is a monotonically increasing functions of $M$ for every value
of $P_s$ and $d$. In addition, the energy consumption of the sensor
and the sink circuitry during $T^{FS}_{ac}$ is computed as
$(\mathcal{P}^{FS}_{ct}+\mathcal{P}^{FS}_{cr})T^{FS}_{ac}$. For the
sensor node with MFSK, we denote the power consumption of frequency
synthesizer, filters and power amplifier as $\mathcal{P}^{FS}_{Sy}$,
$\mathcal{P}^{FS}_{Filt}$ and $\mathcal{P}^{FS}_{Amp}$,
respectively. In this case,
$\mathcal{P}^{FS}_{ct}=\mathcal{P}^{FS}_{Sy}+\mathcal{P}^{FS}_{Filt}+\mathcal{P}^{FS}_{Amp}$.
It is shown that the relationship between $\mathcal{P}^{FS}_{Amp}$
and the transmission power of an MFSK signal is
$\mathcal{P}^{FS}_{Amp}=\alpha^{FS} \mathcal{P}^{FS}_{t}$, where
$\alpha^{FS}$ is determined based on type of power amplifier. For
instance for a class B power amplifier, $\alpha^{FS}=0.33$
\cite{Cui_GoldsmithITWC0905}, \cite{TangITWC0407}. For the power
consumption of the sink circuity, we use the fact that each branch
in NC-MFSK demodulator consists of band pass filters followed by an
envelop detector. Also, we assume that the sink node uses a
Low-Noise Amplifier (LNA) which is generally placed at the front-end
of a RF receiver circuit, an Intermediate-Frequency Amplifier (IFA),
and an ADC unit, regardless of type of deployed pass-band modulation
scheme. Thus, denoting $\mathcal{P}^{FS}_{LNA}$,
$\mathcal{P}^{FS}_{Filr}$, $\mathcal{P}^{FS}_{ED}$,
$\mathcal{P}^{FS}_{IFA}$ and $\mathcal{P}^{FS}_{ADC}$ as the power
consumption of LNA, filters, envelop detector, IF amplifier and ADC,
respectively, the power consumption of the sink circuitry can be
obtained as
$\mathcal{P}^{FS}_{cr}=\mathcal{P}^{FS}_{LNA}+M\times(\mathcal{P}^{FS}_{Filr}+\mathcal{P}^{FS}_{ED})+\mathcal{P}^{FS}_{IFA}+\mathcal{P}^{FS}_{ADC}$.
In addition, it is shown that the power consumption during
transition mode period $T^{FS}_{tr}$ is governed by the frequency
synthesizer \cite{Wang_ISLPED2001}. Taking this into account, the
energy consumption during  $T^{FS}_{tr}$ is obtained as
$\mathcal{P}^{FS}_{tr}T^{FS}_{tr}=1.75
\mathcal{P}^{FS}_{Sy}T^{FS}_{tr}$  \cite{Karl_Book2005}. As a
result, the total energy consumption of a NC-MFSK scheme for
transmitting $N$ bits in each period $T_N$, under the constraint $M
\leq M_{max}$ and for a given $P_s$ is obtained as
\begin{equation}\label{energy_totFSK}
\mathcal{E}^{FS}_N = (1+\alpha^{FS}) \left[\left(
1-(1-P_s)^{\frac{1}{M-1}}\right)^{-1}-2 \right]\dfrac{\mathcal{L}_d
N_0}{\Omega} \dfrac{N}{\log_2 M}+
(\mathcal{P}^{FS}_{c}-\mathcal{P}^{FS}_{Amp})\dfrac{MN}{B\log_{2}M}+1.75
\mathcal{P}^{FS}_{Sy}T^{FS}_{tr}.
\end{equation}

\textbf{M-ary QAM:} For $M$-ary QAM with square constellation, each
$b=\log_{2}M$ bits of the message is mapped to a complex symbol
$S_i$, $i=0,1,...,M-1$, where the constellation size $M$ is a power
of 4. Assuming raised-cosine filter (with a proper filter roll-off)
is used for pulse shaping, the channel bandwidth of MQAM is
determined as $B \approx \frac{1}{2T^{QA}_s}$, where $T^{QA}_s$
represents the MQAM symbol duration. Using the data rate
$R^{QA}=\frac{b}{T^{QA}_s}$ (bits/sec), the bandwidth efficiency of
 MQAM is obtained as $B^{QA}_{eff}\triangleq
\frac{R^{QA}}{B}=2\log_2M~\textrm{(bits/s/Hz)}$. It is observed that
$B^{QA}_{eff}$ is a logarithmically increasing function of $M$. To
address the effect of increasing $M$ on the energy efficiency, we
first derive the relationship between $M$ and active mode duration
$T^{QA}_{ac}$. Recalling that during period $T^{QA}_{s}$, we have
$b$ bits, it is concluded that
\begin{equation}\label{active_QAM}
T^{QA}_{ac}=\dfrac{N}{b}T^{QA}_{s}=\dfrac{N}{2B\log_2M}.
\end{equation}
We can see from (\ref{active_QAM}) that increasing $M$ results in
decreasing in $T^{QA}_{ac}$. Also compared to (\ref{active1}), it is
concluded that $\frac{T^{QA}_{ac}}{T^{FS}_{ac}}=\frac{1}{M} <1$. To
obtain the transmit energy consumption $\mathcal{P}^{QA}_t
T^{QA}_{ac}$, we use the similar arguments as MFSK. The SER
conditioned upon $\gamma_{i}$ of a coherent MQAM is given by
\cite[pp. 226]{Simon2005}
\begin{eqnarray}
P_s(\gamma_{i})&=&4\left(1-\dfrac{1}{\sqrt{M}}\right)  Q\left( \sqrt{\dfrac{3\gamma_i}{M-1}} \right)-4\left(1-\dfrac{1}{\sqrt{M}}\right)^2  Q^2\left( \sqrt{\dfrac{3\gamma_i}{M-1}} \right)\\
&\leq& 4\left(1-\dfrac{1}{\sqrt{M}}\right)  Q\left( \sqrt{\dfrac{3\gamma_i}{M-1}} \right)\\
&\stackrel{(a)}{\leq}& 2\left(1-\dfrac{1}{\sqrt{M}}\right) \exp
\left( -\dfrac{3\gamma_i}{2(M-1)} \right),
\end{eqnarray}
where $Q(x)\triangleq \frac{1}{\sqrt{2
\pi}}\int_{x}^{\infty}e^{-u^2/2}du$ is the area under the tail of
the Gaussian distribution, and $(a)$ comes from $Q(x) \leq
\frac{1}{2} e^{-x^2/2}$, $x>0$. Thus, the average SER of a coherent
MQAM is upper bounded by
\begin{eqnarray}
P_{s}&=&\int_{0}^{\infty}P_{s}(\gamma_{i})f_{\gamma}(\gamma_{i})d\gamma_{i}\leq
\dfrac{2}{\bar{\gamma}^{QA}}\left(1-\dfrac{1}{\sqrt{M}}\right)
\int_{0}^{\infty}
e^{-\frac{3\gamma_i}{2(M-1)}}.e^{-\frac{\gamma_i}{\bar{\gamma}^{QA}}}d\gamma_{i}\\
&=&\dfrac{4(M-1)}{3\bar{\gamma}^{QA}+2(M-1)}\left(1-\dfrac{1}{\sqrt{M}}
\right),
\end{eqnarray}
where $\bar{\gamma}^{QA}
=\frac{\Omega}{\mathcal{L}_d}\frac{\mathcal{E}^{QA}_t}{N_0}$ denotes
the average received SNR with energy per symbol
$\mathcal{E}^{QA}_t$. As a result,
\begin{equation}
\mathcal{E}^{QA}_t \triangleq \mathcal{P}^{QA}_t T^{QA}_s \leq
\dfrac{2(M-1)}{3}\left[ 2 \left( 1-\dfrac{1}{\sqrt{M}}
\right)\dfrac{1}{P_s}-1 \right] \dfrac{\mathcal{L}_dN_0}{\Omega}.
\end{equation}
With a similar argument as MFSK and by approximating the above upper
bound as an equality, the energy consumption of transmitting $N$
bits during active mode period is computed as
\begin{equation}\label{energy_trans_MQAM}
\mathcal{P}^{QA}_t T^{QA}_{ac} =
\dfrac{T^{QA}_{ac}}{T^{QA}_s}\mathcal{E}^{QA}_t \approx
\dfrac{2(M-1)}{3}\left[ 2 \left( 1-\dfrac{1}{\sqrt{M}}
\right)\dfrac{1}{P_s}-1 \right] \dfrac{\mathcal{L}_dN_0}{\Omega}
\dfrac{N}{\log_2 M},
\end{equation}
which is a monotonically increasing functions of $M$ for every value
of $P_s$ and $d$. The energy consumption of the sensor and sink
circuitry during active mode period $T^{QA}_{ac}$ for MQAM scheme is
computed as
$(\mathcal{P}^{QA}_{ct}+\mathcal{P}^{QA}_{cr})T^{QA}_{ac}$.
According to Fig. \ref{fig: MQAM_Mod}, for the sensor node with
MQAM,
$\mathcal{P}^{QA}_{ct}=\mathcal{P}^{QA}_{DAC}+\mathcal{P}^{QA}_{FS}+\mathcal{P}^{QA}_{Mix}+\mathcal{P}^{QA}_{Filt}+\mathcal{P}^{QA}_{Amp}$,
where $\mathcal{P}^{QA}_{DAC}$ and $\mathcal{P}^{QA}_{Mix}$ denote
the power consumption of DAC and mixer. It is shown that the
relationship between $\mathcal{P}^{QA}_{Amp}$ and the transmission
power $\mathcal{P}^{QA}_t$ is given by
$\mathcal{P}^{QA}_{Amp}=\alpha^{QA} \mathcal{P}^{QA}_{t}$, where
$\alpha^{QA}=\frac{\xi}{\vartheta}-1$ with
$\xi=3\frac{\sqrt{M}-1}{\sqrt{M}+1}$ and $\vartheta=0.35$
\cite{Cui_GoldsmithITWC0905}. In addition, the power consumption of
the sink circuitry with coherent MQAM can be obtained as
$\mathcal{P}^{QA}_{cr}=\mathcal{P}^{QA}_{LNA}+\mathcal{P}^{QA}_{Mix}+\mathcal{P}^{QA}_{Sy}+\mathcal{P}^{QA}_{Filr}+\mathcal{P}^{QA}_{IFA}+\mathcal{P}^{QA}_{ADC}$.
Also, with a similar argument as MFSK, we assume that the circuit
power consumption during transition mode period $T^{QA}_{tr}$ is
governed by the frequency synthesizer. As a result, the total energy
consumption of a coherent MQAM system for transmitting $N$ bits in
each period $T_N$ is obtained as
\begin{eqnarray}
\notag \mathcal{E}^{QA}_N &=& (1+\alpha^{QA})
\dfrac{2(M-1)}{3}\left[ 2 \left( 1-\dfrac{1}{\sqrt{M}}
\right)\dfrac{1}{P_s}-1 \right]
\dfrac{\mathcal{L}_{d}N_0}{\Omega} \dfrac{N}{\log_2 M}+\\
\label{energy_totMQAM}&&
(\mathcal{P}^{QA}_{c}-\mathcal{P}^{QA}_{Amp})\dfrac{N}{2B\log_{2}M}
+2 \mathcal{P}^{QA}_{Sy}T^{QA}_{tr}.
\end{eqnarray}

\textbf{Offset-QPSK:} OQPSK referred to as staggered QPSK is used in
IEEE 802.15.4 standard which is the industry standard for WSNs. The
structure of an OQPSK modulator is the same as an QPSK modulator
except that the in-phase (or I-channel) and the quadrature-phase (or
Q-channel) pulse trains are staggered. Since OQPSK differs from QPSK
only by a delay in the Q-channel signal, its error performance on a
linear AWGN channel with ideal coherent detection at the receiver is
the same as that of QPSK \cite{Simon2005}. For this configuration, a
coherent phase reference must be available at the receiver. However,
coherent detection can be costly and increases implementation
complexity due to deriving a reference carrier signal in the
demodulator. On the other hand, since for OQPSK the information is
carried in the phase of the carrier, and noting that non-coherent
receivers are designed to ignore this phase, non-coherent detection
can not be employed with OQPSK modulation. A popular technique which
surpasses utilizing a coherent phase reference is to use
\emph{differential encoding} before classical OQPSK modulator (see
\cite[Fig. 4]{JamshidTech2009}). This is called Differential Offset
QPSK (DOQPSK). In this case, the sensed data stream $(a_1,...,a_N)$
is first differentially encoded twice at the sensor node such that
it is the change from one bit to the next using $\tilde{a}_n=a_n
\oplus \tilde{a}_{n-1}$, where $\oplus$ denotes addition modulo 2.
Then, the encoded bit stream is entered to the classical OQPSK
modulator. For the above configuration, the channel bandwidth and
the data rate are determined by $B \approx \frac{1}{T^{OQ}_s}$ and
$R^{OQ}=\frac{2}{T^{OQ}_s}$ (bits/sec), respectively. As a result,
the bandwidth efficiency of OQPSK is obtained as $B^{OQ}_{eff}
\triangleq \frac{R^{OQ}}{B}=2$ (bits/s/Hz), which is the same as
that of DOQPSK. Since during each symbol period $T^{OQ}_{s}$, we
have $2$ bits, it is concluded that
$T^{OQ}_{ac}=\frac{N}{2}T^{OQ}_{s}=\frac{N}{2B}$. Compared to
(\ref{active1}) and (\ref{active_QAM}), we have $T^{QA}_{ac} <
T^{OQ}_{ac} < T^{FS}_{ac}$. To determine the transmit energy
consumption of differential OQPSK scheme, denoted by
$\mathcal{P}^{OQ}_t T^{OQ}_{ac}$, one would derive
$\mathcal{E}^{OQ}_t$ in terms of SER. The SER conditioned upon
$\gamma_i$ of differential OQPSK for two-bits observation interval
is upper bounded by \cite{SimonITC0603}
\begin{eqnarray}
P_s(\gamma_{i})&\leq& 1-Q_{1}\left(
\sqrt{\dfrac{2+\sqrt{2}}{4}\gamma_{i}},
\sqrt{\dfrac{2-\sqrt{2}}{4}\gamma_{i}}\right)+Q_{1}\left(
\sqrt{\dfrac{2-\sqrt{2}}{4}\gamma_{i}},
\sqrt{\dfrac{2+\sqrt{2}}{4}\gamma_{i}}\right)\\
&\lesssim&\sqrt{\dfrac{1+\sqrt{2}}{2}}\textrm{erfc}\left(
\sqrt{\dfrac{2-\sqrt{2}}{4}\gamma_{i}}\right)\stackrel{(a)}{\leq}
\sqrt{\dfrac{1+\sqrt{2}}{2}}e^{-\frac{2-\sqrt{2}}{4}\gamma_i},
\end{eqnarray}
where $Q_{1}(\alpha,\beta)$ is the first-order Marcum Q-function
\cite{Marcum1950} defined as $Q_{1}(\alpha,\beta) \triangleq
\int_{\beta}^{\infty}xI_0(\alpha x)e^{-\frac{x^2+\alpha^2}{2}}dx$,
and $\textrm{erfc}(x) \triangleq
\frac{2}{\sqrt{\pi}}\int_{x}^{\infty}e^{-u^2}du$ is the
complementary error function. In the above inequalities, $(a)$
follows from $\textrm{erfc}(x) \leq e^{-x^2}$. Thus, the average SER
of differential OQPSK is upper bounded by
\begin{eqnarray}
P_{s}&=&\int_{0}^{\infty}P_{s}(\gamma_{i})f_{\gamma}(\gamma_{i})d\gamma_{i}
\leq
\sqrt{\dfrac{1+\sqrt{2}}{2}}\dfrac{1}{\bar{\gamma}^{OQ}}\int_{0}^{\infty}\exp
\left[ -\left(
\dfrac{2-\sqrt{2}}{4}+\dfrac{1}{\bar{\gamma}^{OQ}}\right)\gamma_i
\right]d\gamma_i\\
&=&\sqrt{\dfrac{1+\sqrt{2}}{2}}\dfrac{4}{(2-\sqrt{2})\bar{\gamma}^{OQ}+4},
\end{eqnarray}
where
$\bar{\gamma}^{OQ}=\frac{\Omega}{\mathcal{L}_d}\frac{\mathcal{E}^{OQ}_t}{N_0}$.
Approximating the above upper bound as an equality, the transmit
energy consumption per each symbol for a given $P_s$ is obtained as
$\mathcal{E}^{OQ}_t \triangleq \mathcal{P}^{OQ}_t T^{OQ}_s \approx
\left[\frac{1}{2-\sqrt{2}} \left(
\frac{4}{P_s}\sqrt{\frac{1+\sqrt{2}}{2}}-4 \right)
\right]\frac{\mathcal{L}_dN_0}{\Omega}$. Thus, the energy
consumption of transmitting $N$ bits during active mode is computed
as
\begin{equation}\label{energy_trans_OQPSK}
\mathcal{P}^{OQ}_t T^{OQ}_{ac} =
\dfrac{T^{OQ}_{ac}}{T^{OQ}_s}\mathcal{E}^{OQ}_t \approx
\left[\dfrac{1}{2-\sqrt{2}} \left(
\dfrac{4}{P_s}\sqrt{\dfrac{1+\sqrt{2}}{2}}-4 \right)
\right]\dfrac{\mathcal{L}_dN_0}{\Omega}\dfrac{N}{2}.
\end{equation}

The energy consumption of the sensor and the sink circuitry during
active mode period $T^{OQ}_{ac}$ for differential OQPSK is computed
as $(\mathcal{P}^{OQ}_{ct}+\mathcal{P}^{OQ}_{cr})T^{OQ}_{ac}$. With
a similar argument, for the sensor node with differential OQPSK,
$\mathcal{P}^{OQ}_{ct} \approx
\mathcal{P}^{OQ}_{DAC}+\mathcal{P}^{OQ}_{FS}+\mathcal{P}^{OQ}_{Mix}+\mathcal{P}^{OQ}_{Filt}+\mathcal{P}^{OQ}_{Amp}$,
where we assume that the power consumption of differential encoder
blocks are negligible, and $\mathcal{P}^{OQ}_{Amp}=\alpha^{OQ}
\mathcal{P}^{OQ}_{t}$, with $\alpha^{OQ}=0.33$. In addition, the
power consumption of the sink circuitry with differential detection
OQPSK can be obtained as
$\mathcal{P}^{OQ}_{cr}=\mathcal{P}^{OQ}_{LNA}+\mathcal{P}^{OQ}_{Mix}+\mathcal{P}^{OQ}_{FS}+\mathcal{P}^{OQ}_{Filr}+\mathcal{P}^{OQ}_{IF}+\mathcal{P}^{OQ}_{ADC}$.
As a result, the total energy consumption of a differential OQPSK
system for transmitting $N$ bits in each period $T_N$ is obtained as
\begin{equation}\label{energy_totOQPSK}
\mathcal{E}^{OQ}_N = (1+\alpha^{OQ}) \left[\dfrac{1}{2-\sqrt{2}}
\left( \dfrac{4}{P_s}\sqrt{\dfrac{1+\sqrt{2}}{2}}-4 \right)
\right]\dfrac{\mathcal{L}_dN_0}{\Omega}\dfrac{N}{2}
+(\mathcal{P}^{OQ}_{c}-\mathcal{P}^{OQ}_{Amp})\dfrac{N}{2B}+2
\mathcal{P}^{OQ}_{Sy}T^{OQ}_{tr}.
\end{equation}

\section{Energy Consumption Analysis of UWB Modulation Schemes}\label{Analysis_Ch 4}
Ultra-wideband is a very short-range wireless technology for
transmitting information over a wide spectrum at least 500 MHZ or
greater than 20\% of the center frequency, whichever is less,
according to Federal Communications Commission (FCC)
\cite{Oppermann2004}. For instance, a UWB signal centered at 2.4 GHz
would have a minimum bandwidth of 500 MHz. The main advantages of
UWB modulation schemes are their immunity to multipath, very low
transmission power and simple transceiver circuity. Among current
UWB receivers, the energy detection based non-coherent receiver is
the simplest and most practical one to implement, bypassing coherent
phase reference requirement at the receiver. These inherent
advantages make the UWB technology an attractive candidate for using
in low data rate and ultra-low power wireless sensor networking
applications, e.g., positioning, monitoring and control. In the
following, we investigate OOK and M-ary PPM modulation schemes from
energy and bandwidth efficiency points of view\footnote{For
simplicity of notations, we use the superscripts `OK' and `PP' for
OOK and M-ary PPM modulation schemes, respectively.}.

\textbf{On-Off Keying:} Similar to the pass-band modulation schemes,
we assume that the sensor node aims to transmit $(a_1,...,a_N)$
during $T_N$ period using an OOK modulation scheme. Note that for
OOK, $b=\log_2 M=1$. The structure of an OOK-based transmitter is
very simple as depicted in Fig. \ref{fig: OOK_TX}. The UWB pulse
generator is followed by an OOK modulator which is controlled by
$(a_1,...,a_N)$. For this configuration, an OOK transmitted signal
correspond to bit $a_i$ is given by
$x^{OK}_i(t)=\sqrt{\mathcal{E}^{OK}_t}a_i p(t-iT^{OK}_s)$, where
$p(t)$ is the radiated ultra-short pulse of width $T^{OK}_p$ with
unit energy, $\mathcal{E}^{OK}_t$ is the transmit energy consumption
per each symbol or bit, and $T_{s}^{OK}$ is the OOK symbol duration.
A typical pulse $p(t)$ which is widely used in UWB systems is the
ultra-short \emph{Gaussian monocycle} with duration $T_p$ (Fig.
\ref{fig: OOK_TX}). This monocycle is a wideband signal with
bandwidth approximately equal to $\frac{1}{T_p}$ \footnote{Since it
is assumed that $B \approx \frac{1}{T_p}$ is fixed, we use the same
$T_p$ for all kind of UWB modulation schemes and drop superscript
for $T_p$.}. In time domain, a Gaussian monocycle is derived by the
first derivative of the Gaussian function
$\mathfrak{T}(t)=6A_c\sqrt{\frac{e\pi}{3}}\frac{t}{T_p}e^{-6 \pi
\left( \frac{t}{T_p} \right)^2}$, where $A_c$ is the peak amplitude
of the monocycle \cite [p.108]{Fazel_Book2003}.

The ratio $\frac{T_{p}}{T^{OK}_s}$ is defined as the
\emph{duty-cycle factor} of an OOK signal, which is the fractional
on-time of the OOK ``1'' pulse. In addition, the channel bandwidth
and the data rate of an OOK are determined as $B \approx
\frac{1}{T_{p}}$ and $R^{OK}=\frac{1}{T_{s}^{OK}}$ (bits/sec),
respectively. As a result, the bandwidth efficiency of an OOK are
obtained as $B^{OK}_{eff} \triangleq
\frac{R^{OK}}{B}=\frac{T_p}{T^{OK}_s} \leq 1$ (bits/s/Hz). Note that
during transmission bit $a_i=0$, the filter and the power amplifier
of the OOK modulator are turn off. However, it does not mean that
the receiver is turn off as well. For this reason, we still use the
same definition for active mode period $T^{OK}_{ac}$ as pass-band
schemes, and that is given by $T^{OK}_{ac}=\frac{N}{b}T^{OK}_{s}=N
T^{OK}_{s}$. Depend upon the duty-cycle factor, $T_{ac}^{OK}$ can be
expressed in terms of bandwidth $B$. For instance, for an OOK with
duty-cycle factor $\frac{T_{p}}{T^{OK}_s}=\frac{1}{2}$, we have
$T^{OK}_{ac}=2N T_{p}=\frac{2N}{B}$, and $B^{OK}_{eff}=\frac{1}{2}$.

For energy consumption analysis, we describe a conventional
non-coherent receiver relating to the OOK scheme which is an energy
detector receiver \cite{Urkowitz1967, Paquelet2004}. This structure
consists of a filter on the considered band, a square-law block, and
an integrator followed by a decision block with an optimum threshold
level (see \cite[Fig. 7]{JamshidTech2009}). In addition, a low-noise
amplifier is included in front-end. Using this non-coherent
receiver, no oscillator is required for phase synchronization, and
the receiver can turn on quickly. Note that when bit ``0'' is
transmitted using OOK, the sensor node is silent, while at sink
node, only band-pass noise is presented to the envelope detector. It
can be shown in \cite[pp. 490-504]{Couch_Book2001} that the BER
conditioned upon $\gamma_{i}$ of a non-coherent OOK, denoted by
$P_b(\gamma_{i})$, is upper bounded by $P_b(\gamma_{i}) \leq
\frac{1}{2} e^{-\frac{\gamma_{i}}{2}}$. Thus, the average BER of an
OOK with non-coherent receiver, denoted by $P_b$, is upper bounded
by
\begin{eqnarray}
P_{b}&=&\int_{0}^{\infty}P_{b}(\gamma_{i})f_{\gamma}(\gamma_{i})d\gamma_{i} \leq \dfrac{1}{2\bar{\gamma}^{OK}}\int_{0}^{\infty} e^{-\gamma_{i}\left(\frac{1}{2}+\frac{1}{\bar{\gamma}^{OK}} \right)}d\gamma_i\\
&=&\dfrac{1}{\bar{\gamma}^{OK}+2},
\end{eqnarray}
where
$\bar{\gamma}^{OK}=\frac{\Omega}{\mathcal{L}_d}\frac{\mathcal{E}^{OK}_t}{N_0}$
denotes the average received SNR. It should be noted that the
average SER of OOK is the same as average BER. By approximating the
above upper bound as an equality, the transmit energy consumption
per each symbol for a given $P_b$ is obtained as $\mathcal{E}^{OK}_t
\triangleq \mathcal{P}^{OK}_t T_p \approx \left(\frac{1}{P_b}-2
\right)\frac{\mathcal{L}_dN_0}{\Omega}$, which is correspond to
transmitting OOK bit ``1''. Note that the energy consumption of
transmitting $N$ bits during active mode (i.e., $\mathcal{P}^{OK}_t
T^{OK}_{ac}$) is equivalent to the energy consumption of
transmitting $L$ bits ``1'' in $(a_1,...a_N)$, where $L$ is a
binomial random variable with parameters $(N,q)$. Assuming
uncorrelated and equally likely binary data $a_i$, we have
$q=\frac{1}{2}$. Hence, $\mathcal{P}^{OK}_t T^{OK}_{ac} =L
\mathcal{E}^{OK}_t \approx L \left(\frac{1}{P_b}-2
\right)\frac{\mathcal{L}_dN_0}{\Omega}$, where $L$ has the
probability mass function $\textrm{Pr}\{ L=\ell
\}=\binom{N}{\ell}\left(\frac{1}{2}\right)^N$ with
$\mathbb{E}[L]=\frac{N}{2}$.

The energy consumption of the sensor and sink circuitry during
$T^{OK}_{ac}$ for a non-coherent OOK is computed as
$(\mathcal{P}^{OK}_{ct}+\mathcal{P}^{OK}_{cr})T^{OK}_{ac}$. We
denote the power consumptions of pulse generator, power amplifier
and filter as $\mathcal{P}^{OK}_{PG}$, $\mathcal{P}^{OK}_{Amp}$ and
$\mathcal{P}^{OK}_{Filt}$, respectively. Hence, the energy
consumption of the sensor node circuitry during $T^{OK}_{ac}$ is
represented as a function of random variable $L$ as
$\mathcal{P}^{OK}_{ct}T^{OK}_{ac} \approx
\mathcal{P}^{OK}_{PG}T^{OK}_{ac}+ L T_p
\left(\mathcal{P}^{OK}_{Filt}+\mathcal{P}^{OK}_{Amp}\right)$, where
the factor $L T_p $ comes from the fact that filter and power
amplifier are active only during transmitting $L$ bits ``1''. We
assume that $\mathcal{P}^{OK}_{Amp}=\alpha^{OK}
\mathcal{P}^{OK}_{t}$, with $\alpha^{OK}=0.33$. In addition, the
energy consumption of the sink circuitry with non-coherent detection
OOK can be obtained as
$\mathcal{P}^{OK}_{cr}T^{OK}_{ac}=\left(\mathcal{P}^{OK}_{LNA}+\mathcal{P}^{OK}_{ED}+\mathcal{P}^{OK}_{Filr}+\mathcal{P}^{OK}_{Int}+\mathcal{P}^{OK}_{ADC}\right)
T^{OK}_{ac}$, where $\mathcal{P}^{OK}_{Int}$ is the power
consumption of integrator unit. Also, with a similar argument as
pass-band modulations, we assume that the circuit power consumption
during $T^{OK}_{tr}$ is governed by the pulse generator block. As a
result, the total energy consumption of a non-coherent OOK for
transmitting $N$ bits is obtained as a function of random variable
$L$ as follows:
\begin{equation}\label{energy_totOOK1}
\mathcal{E}^{OK}_N (L)=  (1+\alpha^{OK})L \left(\dfrac{1}{P_b}-2
\right)\dfrac{\mathcal{L}_dN_0}{\Omega}
+(\mathcal{P}^{OK}_{cr}+\mathcal{P}^{OK}_{PG})\dfrac{2N}{B}+\dfrac{L}{B}\mathcal{P}^{OK}_{Filt}+2
\mathcal{P}^{OK}_{PG}T^{OK}_{tr},
\end{equation}
where we use $T^{OK}_{ac}=\frac{2N}{B}$. Using
$\mathbb{E}[L]=\frac{N}{2}$, the average $\mathcal{E}^{OK}_N (L)$ is
computed as
\begin{equation}\label{energy_totOOK01}
\mathcal{E}^{OK}_N \triangleq \mathbb{E}\left[\mathcal{E}^{OK}_N
(L)\right]= (1+\alpha^{OK}) \left(\dfrac{1}{P_b}-2
\right)\dfrac{\mathcal{L}_dN_0}{\Omega}\dfrac{N}{2}
+(\mathcal{P}^{OK}_{cr}+\mathcal{P}^{OK}_{PG})\dfrac{2N}{B}+\dfrac{N}{2B}\mathcal{P}^{OK}_{Filt}+2
\mathcal{P}^{OK}_{PG}T^{OK}_{tr}.
\end{equation}

\textbf{M-ary PPM:} In an M-PPM modulator each $b \triangleq \log_2
M$ bits are encoded by transmitting a single pulse in one of $M=2^b$
possible time-shifts. This process is repeated every $T^{PP}_{s}$
seconds. An M-PPM signal constellation consists of a set of $M$
orthogonal pulses (in time) with equal energy. This is the
time-domain dual to MFSK which uses a set of $M$ orthogonal
carriers. Assuming Gaussian monocycle pulse with duration $T_p$ for
pulse shaping, the channel bandwidth and data rate of M-PPM are
determined as $B \approx \frac{1}{T_p}$ and
$R^{PP}=\frac{b}{T^{PP}_s}$, respectively, where $T^{PP}_s=M T_p$
represents the time duration of a symbol. As a result, the bandwidth
efficiency of an M-PPM scheme is obtained as $B^{PP}_{eff}
\triangleq \frac{R^{PP}}{B}=\frac{\log_2
M}{M}~\textrm{(bits/s/Hz)}$. For $M=2$, the bandwidth efficiency of
M-PPM is the same as that of OOK with $50\%$ duty-cycle. For other
values of $M$, the same $B^{PP}_{eff}$ is achieved through
$\frac{T_p}{T^{OK}_s}=\frac{\log_2 M}{M}$ with adjusting $T^{OK}_s$.
In addition, similar to MFSK scheme, it can be seen that increasing
constellation size $M$ leads to decrease in the bandwidth efficiency
in M-PPM. To address the impact of increasing $M$ on energy
efficiency, let derive the relationship between $M$ and the active
mode duration $T^{PP}_{ac}$. Noting that during each symbol period
$T^{PP}_{s}$, we have $b$ bits, it is concluded that
$T^{PP}_{ac}=\frac{N}{b}T^{PP}_{s}=\frac{MN}{B\log_2 M}$. With a
similar argument as MFSK, increasing $M$ results in increasing in
$T^{PP}_{ac}$, however $T^{PP}_{ac}$ is upper bounded by
$T_N-T^{PP}_{tr}$. Thus, the maximum constellation size $M$, denoted
by $M_{max}\triangleq 2^{b_{max}}$, for M-PPM is calculated by
$\frac{2^{b_{max}}}{b_{max}}=\frac{B}{N}(T_{N}-T^{PP}_{tr})$.

As mentioned before, among current UWB receivers, the energy
detection based non-coherent receiver is the simplest and most
practical one to implement. The optimum non-coherent M-PPM receiver
consists of a bank of $M$ matched filters, each followed by an
envelop detector (see Fig. B.13 in \cite{Xiong_2006}). At the
sampling times $t=\ell T^{PP}_s$, a maximum-likelihood makes
decision based on the largest filter output. It is shown that for
non-coherent M-PPM, the SER conditioned upon $\gamma_{i}$ is
obtained as follows \cite[p. 969]{Xiong_2006}
\begin{equation}\label{condition_gamma}
P_{s}(\gamma_{i})=\sum_{k=1}^{M-1}\dfrac{(-1)^{k+1}}{k+1}\binom{M-1}{k}e^{-\frac{k}{k+1}\gamma_{i}}.
\end{equation}
The leading term of (\ref{condition_gamma}) provides an upper bound
as $P_{s}(\gamma_{i}) \leq \frac{M-1}{2}e^{-\frac{\gamma_{i}}{2}}$.
For $M=2$, this upper bound becomes the exact expression. The
average SER of a non-coherent M-PPM is then upper bounded by
\begin{eqnarray}
P_{s}&=&\int_{0}^{\infty}P_{s}(\gamma_{i})f_{\gamma}(\gamma_{i})d\gamma_{i}
\leq \dfrac{M-1}{2\bar{\gamma}^{PP}}\int_{0}^{\infty} e^{-\gamma_{i}\left(\frac{1}{2}+\frac{1}{\bar{\gamma}^{PP}} \right)}d\gamma_i\\
\label{upper_PPM}&=&\dfrac{M-1}{\bar{\gamma}^{PP}+2},
\end{eqnarray}
where $\bar{\gamma}^{PP}
=\frac{\Omega}{\mathcal{L}_d}\frac{\mathcal{E}^{PP}_t}{N_0}$. For
large $\bar{\gamma}^{PP}$, the upper bound (\ref{upper_PPM}) is the
same as the upper bounded $P_s$ for MFSK. By approximating the upper
bound (\ref{upper_PPM}) as an equality, the transmit energy
consumption per each symbol for a given $P_s$ is obtained as
$\mathcal{E}^{PP}_t \triangleq \mathcal{P}^{PP}_t T_p \approx
\left(\frac{M-1}{P_s}-2 \right)\frac{\mathcal{L}_dN_0}{\Omega}$.
Noting that during active mode period $T_{ac}^{PP}$, we have
$\frac{N}{b}$ monocycle pulses, the energy consumption of
transmitting $N$ bits during active mode period is computed as
$\mathcal{P}^{PP}_t T^{PP}_{ac} =
 \frac{N}{\log_2 M}\mathcal{E}^{PP}_t \approx
\left(\frac{M-1}{P_s}-2
\right)\frac{\mathcal{L}_dN_0}{\Omega}\frac{N}{\log_2 M}$. It can be
seen that the energy consumption of transmitting $N$ bits during
active mode period for M-PPM scheme is a monotonically increasing
function of $M$.

The energy consumption of the sensor and the sink circuitry during
active mode period $T^{PP}_{ac}$ for a non-coherent M-PPM is
computed as
$(\mathcal{P}^{PP}_{ct}+\mathcal{P}^{PP}_{cr})T^{PP}_{ac}$. The
energy consumption of the sensor node circuitry during active mode
duration is given by $\mathcal{P}^{PP}_{ct}T^{PP}_{ac} \approx
\left(\mathcal{P}^{PP}_{PG}+
\mathcal{P}^{PP}_{Filt}+\mathcal{P}^{PP}_{Amp}\right)\frac{N}{B\log_2
M}$, where the factor $\frac{N}{B\log_2 M}$ comes from the fact that
transmitter circuitry is active only during $T_{p}$ in each period
$T^{PP}_{s}$. We assume that $\mathcal{P}^{PP}_{Amp}=\alpha^{PP}
\mathcal{P}^{PP}_{t}$, with $\alpha^{PP}=0.33$. With a similar
argument as MFSK, the energy consumption of the sink circuitry with
non-coherent detection M-PPM can be obtained as
$\mathcal{P}^{PP}_{cr}T^{PP}_{ac}=\left(\mathcal{P}^{PP}_{LNA}+M\times\left(\mathcal{P}^{PP}_{ED}+\mathcal{P}^{PP}_{Filr}\right)+\mathcal{P}^{PP}_{ADC}\right)
\frac{MN}{B\log_2 M}$. As a result, the total energy consumption of
a non-coherent M-PPM system for transmitting $N$ bits is obtained as
\begin{eqnarray}
\mathcal{E}^{PP}_N &=&  (1+\alpha^{PP}) \left(\dfrac{M-1}{P_s}-2 \right)\dfrac{\mathcal{L}_dN_0}{\Omega}\dfrac{N}{\log_2 M}
+(\mathcal{P}^{PP}_{ct}-\mathcal{P}^{PP}_{Amp})\dfrac{N}{B\log_2 M}+\\
\label{energy_totOOK}&&\mathcal{P}^{PP}_{cr}\dfrac{MN}{B\log_2 M}+2
\mathcal{P}^{PP}_{PG}T^{PP}_{tr}.
\end{eqnarray}

\section{Numerical Results}\label{simulation_Ch5}
In this section, we present some numerical evaluations using
realistic parameters in IEEE 802.15.4 standard and present
state-of-the art technology to confirm the energy efficiency
analysis of the pass-band and UWB modulation schemes mentioned in
Sections \ref{Analysis_Ch 3} and \ref{Analysis_Ch 4}. We also
investigate the impact of constellation size $M$ and distance $d$ on
the energy consumptions for M-ary modulation schemes. For numerical
results, we assume that all modulation schemes operate in $f_0=$2.4
GHz Industrial Scientist and Medical (ISM) unlicensed band utilized
in IEEE 802.15.14 for WSNs \cite{IEEE_802_15_4_2006}. Also according
to FCC 15.247 RSS-210 standard for United States/Canada, the maximum
allowed antenna gain is 6 dBi \cite{FreeScale2007}. In this work, we
assume that $\mathcal{G}_t=\mathcal{G}_r=5$ dBi. Thus for $f_0=$2.4
GHz, $\mathcal{L}_1~ \textrm{(dB)}\triangleq
10\log_{10}\left(\frac{(4 \pi)^2}{\mathcal{G}_t \mathcal{G}_r
\lambda^2}\right) \approx 30~ \textrm{dB}$, where $\lambda
\triangleq \frac{3\times 10^8}{f_0}$ in meter. As mentioned before,
the channel bandwidth of UWB schemes is much wider than that of
pass-band ones. Thus, due to the dependency of total energy
consumption on bandwidth, comparing the energy efficiency of UWB
schemes with that of pass-band ones would not meaningful. For this
reason, we evaluate the energy efficiency of each category
separately.

\textbf{Pass-Band Modulation:} We assume that in each period $T_N$,
the sensed data frame size $N=1024$ bytes (or equivalently $N=8192$
bits) is generated for transmission for all the pass-band modulation
schemes, where $T_N$ is assumed to be 1.4 second. The channel
bandwidth is assumed to be $B=62.5$ KHz, according to IEEE 802.15.4
standard \cite[p. 49]{IEEE_802_15_4_2006}. The power consumption of
LNA and IF amplifier are considered about 9 mw
\cite{Bevilacqua_IJSSC1204} and 3 mw \cite{Cui_GoldsmithITWC0905,
TangITWC0407}, respectively, which are assumed to be the same for
all pass-band modulations. The power consumption of frequency
synthesizer in MFSK is supposed to be 10 mw \cite{Wang_ISLPED2001}.
We use this value for the sinusoidal carrier generation in MQAM and
OQPSK as well. We consider 7 mw power consumption for ADC and DAC
\cite{Bravos_PIMRC2005}. In addition, we assume that the
transceivers of the pass-band modulations are able to achieve values
for $T_{tr}$ of about 5 $\mu s$ for MFSK and 20 $\mu s$ for MQAM and
OQPSK \cite{Cui_GoldsmithITWC0905, Bravos_PIMRC2005}. Table I shows
the system parameters for simulation.

It is concluded from $ \frac{2^{b_{max}}}{\zeta
b_{max}}=\frac{B}{N}(T_{N}-T^{FS}_{tr})$ that $M_{max} \approx 64$
(or equivalently $b_{max} \approx 6$) for NC-MFSK. Since, there is
no constraint for $M$ in MQAM, we choose the range $2 \leq M \leq
64$ for MQAM as well to be consistent with MFSK. Recalling from
(\ref{energy_totFSK}), the total energy consumption
$\mathcal{E}_{N}^{FS}$ is a monotonically increasing function of $M$
for different values of $d$. While, the total energy consumption
$\mathcal{E}_{N}^{QA}$ of MQAM in terms of $M$ displays different
trend for each distance $d$ as depicted in Fig. \ref{fig:
Total_Energy}. For instance, for $d=50$ m, the optimum value for $M$
that minimizes $\mathcal{E}_{N}^{QA}$ is 4 (equivalent to 4QAM
scheme). Although the second term of (\ref{energy_totMQAM})
correspond to the circuit energy consumption of MQAM is a
monotonically decreasing function of $M$, however for large values
of $d$, the total energy consumption is governed by the
$\mathcal{P}^{QA}_{t}T_{ac}^{QA}$ which results in
$\mathcal{E}_{N}^{QA}$ to be a monotonically increasing function of
$M$ when $d$ increases. Also, Fig. \ref{fig: Total_Energy_dm}
compares the energy efficiency of the aforemention pass-band
modulations versus $M$ for $P_s=10^{-3}$ and different values of
$d$. It is revealed from Fig. \ref{fig: Total_Energy_dm}-a that for
$M < 35$, NC-MFSK is more energy efficient than MQAM, OQPSK and
coherent MFSK for $d=10$ m. In addition, NC-MFSK for small $M$
benefits from low complexity and cost in implementation over other
schemes. Furthermore, as shown in Fig. \ref{fig: Total_Energy_dm}-b,
the total energy consumption of both MFSK and MQAM for large $d$
increase logarithmically when $M$ increases. Also, it can be seen
that non-coherent 4FSK scheme consumes much less energy than the
other schemes in short ranges of $d$.

Up to know, we have investigated the energy efficiency of pass-band
modulation schemes under the assumption of Rayleigh fading channel
model with path-loss, where there is no Line-Of-Sight (LOS) path
between sensor and sink nodes. Of course, the Rayleigh fading
channel is known to be a reasonable model for fading encountered in
many wireless networks. However, it is also of interest to evaluate
the energy efficiency of pass-band modulation schemes operating over
the more general Rician model which includes a strong direct LOS
path. This is because when WSNs work in short range environments,
there is a high probability of LOS propagation \cite{WyneITWC0809}.
For this purpose, let assume that the instantaneous channel
coefficient correspond to symbol $i$ is
$G_i=\frac{h_i}{\mathcal{L}_d}$, where $h_i$ is assumed to be Rician
distributed with pdf $f_{h_i}(r)=\frac{r}{\sigma^2}e^{-\frac{r^2 +
A^2}{2 \sigma^2} }I_{0} \left( \frac{rA}{\sigma^2} \right),~r \geq
0$, where $A$ denotes the peak amplitude of the dominant signal,
$2\sigma^2 \triangleq \Omega$ is  the average power of non-LOS
multipath components \cite[p. 78]{Goldsmith_Book2005}. For this
model, $\mathbb{E} [\vert h_i \vert ^2]=A^2 +2 \sigma^2 =2 \sigma^2
(1+K)$, where $K(\textrm{dB}) \triangleq 10 \log \frac{A^2}{2
\sigma^2}$ is the Rician factor. The value of $K$ is a measure of
the severity of the fading. For instance, $K(\textrm{dB}) \to
-\infty$ implies Rayleigh fading and $K(\textrm{dB}) \to \infty$
represents AWGN channel. It is seen from Table II that NC-MFSK with
small order of $M$ has minimum total energy consumption compared to
the other schemes in Rician fading channel model with path-loss.
Although MQAM with large $M$ has less energy consumption than
NC-MFSK with the same size $M$ for different values of $K$, however
it is greater than that of NC-4FSK.

The above results make NC-MFSK attractive for using in WSNs, in
particular for short to moderate range applications, since NC-MFSK
already has the advantage of less complexity and cost in
implementation than MQAM, differential OQPSK and coherent MFSK, and
has less total energy consumption. In addition, since for typical
energy-constrained WSNs, data rates are usually low, using small
order of M-ary NC-FSK schemes are desirable. The sacrifice, however,
is the bandwidth efficiency of NC-MFSK (when $M$ increases) which is
a critical factor in band-limited WSNs. But in unlicensed bands
where large bandwidth is available, NC-MFSK can be considered as a
realistic option in WSNs. To have more insight into the behavior of
this scheme, we plot the total energy consumption of NC-MFSK as a
function of $B_{eff}^{FS}$ for different values of $M$ and $d$ in
Fig. \ref{fig: Energy_vs_Bandwidth}. In all cases, we observe that
the minimum $\mathcal{E}_{N}^{FS}$ is achieved at low values of
distance $d$ and for $M=2$, which corresponds to the maximum
bandwidth efficiency $B_{eff}^{FS}=0.5$.

\textbf{UWB Modulation:} In WSNs with very short-range scenarios, it
is desirable to use UWB schemes rather than pass-band ones, due to
the inherent advantages of UWB wireless technology mentioned before.
We assume that total number of transmission bits during $T_{N}$ for
UWB modulations is $N=20000$, where $T_N$ is assumed to be 100 ms.
As previously mentioned, the channel bandwidth of UWB modulation
schemes is $B=500$ MHz for the ISM unlicensed band 2.4 GHz. In
addition, since the main application of UWB systems is in very
short-range scenarios, we assume $1 \leq d \leq 10$ m which is a
feasible range for UWB applications. The power consumption of LNA
for an arbitrarily UWB receiver is considered 3.1 mw
\cite{LeeIEICE0907}. The power consumption of pulse generator is
supposed to be 675 $\mu$w \cite{Zhang2005}. In addition, we assume
that the transceivers of UWB schemes are able to achieve values for
$T_{tr}$ of about 2 ns \cite{Heyn2007}.

Fig. \ref{fig: Total_EnergyPPM_vs_M} plots the total energy
consumption of M-PPM scheme versus $M$ for $P_s=10^{-3}$ and
different distance $d$. The results show that the total energy
consumption $\mathcal{E}_{N}^{PP}$ is still a monotonically
increasing function of $M$ for different values of $d$. Also, Fig.
\ref{fig: Total_Energy_UWB_vs_d} shows the graphs of total energy
consumption of OOK and M-PPM for M=2, 4, 8 in terms of distance $d$
for $P_s=10^{-3}$. As shown in this figure, OOK performs better than
M-PPM from the energy consumption points of view, for $1 \leq d \leq
10$ m. As a result, OOK would be considered an appropriate option
for using in very short-range WSN applications, since OOK already
has the advantage of less complexity and cost in implementation than
M-PPM, and has less total energy consumption.

\section{Conclusion}\label{conclusion_Ch6}
In this paper, we have analyzed in-dept the energy efficiency of
various modulation schemes to find the best scheme in a WSN over
Rayleigh and Rician flat-fading channels with path-loss, using
realistic models and parameters in IEEE 802.15.4 standard.
Experimental results show that NC-MFSK is attractive for using in
WSNs with short to moderate range applications, since NC-MFSK
already has the advantage of less complexity and cost in
implementation than MQAM and differential OQPSK, and has less total
energy consumption. In addition, MFSK has faster start-up time than
other pass-band schemes. Furthermore, since for typical
energy-constrained WSNs, data transmission rates are usually low,
using small order M-ary NC-FSK schemes are desirable. The sacrifice,
however, is the bandwidth efficiency of NC-MFSK when $M$ increases.
However, since most of WSN applications with band-pass modulation
requires low to moderate bandwidth, a loss in the bandwidth
efficiency can be tolerable, in particular for the unlicensed band
applications where large bandwidth is available. In addition, one
would use Minimum Shift Keying (MSK) scheme (as a special case of
FSK), since it has some excellent spectral properties which make it
an attractive alternative when other channel constraints require
bandwidth efficiencies below 1.0 bit/s/Hz. In very short-range WSN
applications, on the other hand, it is desirable to use UWB
modulation schemes, rather than pass-band ones due to ultra-low
power consumption and very low complexity in implementation. It was
shown that OOK has the advantage of very simple complexity and lower
cost in implementation than M-PPM, and has less total energy
consumption.

\appendices
\section{Energy Efficiency of Coherent MFSK }\label{append01}
\renewcommand{\theequation}{A-\arabic{equation}}
\setcounter{equation}{0}

For the coherent MFSK, the precise frequency and carrier phase
reference are available at the sink node. It is shown that the SER
conditioned upon $\gamma_{i}$ of a coherent MFSK is obtained as
 $P_{s}(\gamma_{i})=1-\frac{1}{\sqrt{2
\pi}}\int_{-\infty}^{\infty}
e^{-\frac{u^2}{2}}\left[1-\textrm{erfc}\left(
u+\sqrt{2\gamma_i}\right) \right]^{M-1}du \leq
(M-1)Q(\sqrt{\gamma_i})\leq \frac{M-1}{2}e^{-\frac{\gamma_i}{2}}$,
where we use $Q(x) \leq \frac{1}{2} e^{-\frac{x^2}{2}}$ \cite[p.
383]{Smith2004}. Thus, the average SER of a coherent MFSK is upper
bounded by $P_{s} =
\int_{0}^{\infty}P_{s}(\gamma_{i})f_{\gamma}(\gamma_{i})d\gamma_{i}
\leq \frac{M-1}{2\bar{\gamma}} \int_{0}^{\infty}
e^{-\gamma_i(\frac{1}{2}+\frac{1}{\bar{\gamma}^{FS}})}d\gamma_{i}=\frac{M-1}{\bar{\gamma}^{FS}+2}$,
whit $\bar{\gamma}^{FS}
=\frac{\Omega}{\mathcal{L}_d}\frac{\mathcal{E}^{FS}_t}{N_0}$. As a
result, $\mathcal{E}^{FS}_t \triangleq \mathcal{P}^{FS}_t T^{FS}_s
\leq \left[ \frac{M-1}{P_s}-2 \right]
\frac{\mathcal{L}_dN_0}{\Omega}$. Approximating the above upper
bound as an equality, the energy consumption of transmitting $N$
bits during active mode period is computed as $\mathcal{P}^{FS}_t
T^{FS}_{ac} \approx \left[ \frac{M-1}{P_s}-2 \right]
\frac{\mathcal{L}_dN_0}{\Omega} \frac{N}{\log_2 M}$. The energy
consumption of the sensor circuity for the coherent FSK is the same
as that of the non-coherent case. For the power consumption of the
sink circuity, however, we use the fact that each branch in coherent
MFSK demodulator consists of carrier generator, mixer and band pass
filter. Thus,
$\mathcal{P}^{FS}_{cr}=\mathcal{P}^{FS}_{LNA}+M\times(\mathcal{P}^{FS}_{Sy}+\mathcal{P}^{FS}_{Mix}+\mathcal{P}^{FS}_{Filr})+\mathcal{P}^{FS}_{IFA}+\mathcal{P}^{FS}_{ADC}$.
Using (\ref{active1}) with $\zeta=2$ for coherent MFSK, the total
energy consumption of a coherent MFSK for transmitting $N$ bits
during $T_N$ is obtained as $\mathcal{E}^{FS}_N = (1+\alpha^{FS})
\left[\frac{M-1}{P_s}-2 \right]\frac{\mathcal{L}_d N_0}{\Omega}
\frac{N}{\log_2 M}+
(\mathcal{P}^{FS}_{c}-\mathcal{P}^{FS}_{Amp})\frac{MN}{2B\log_{2}M}+1.75
\mathcal{P}^{FS}_{Sy}T^{FS}_{tr}$, under the constraint $M \leq
M_{max}$.

%\bibliographystyle{IEEE}
%\bibliography{keylatex}

\begin{figure}[t]
\centerline{\psfig{figure=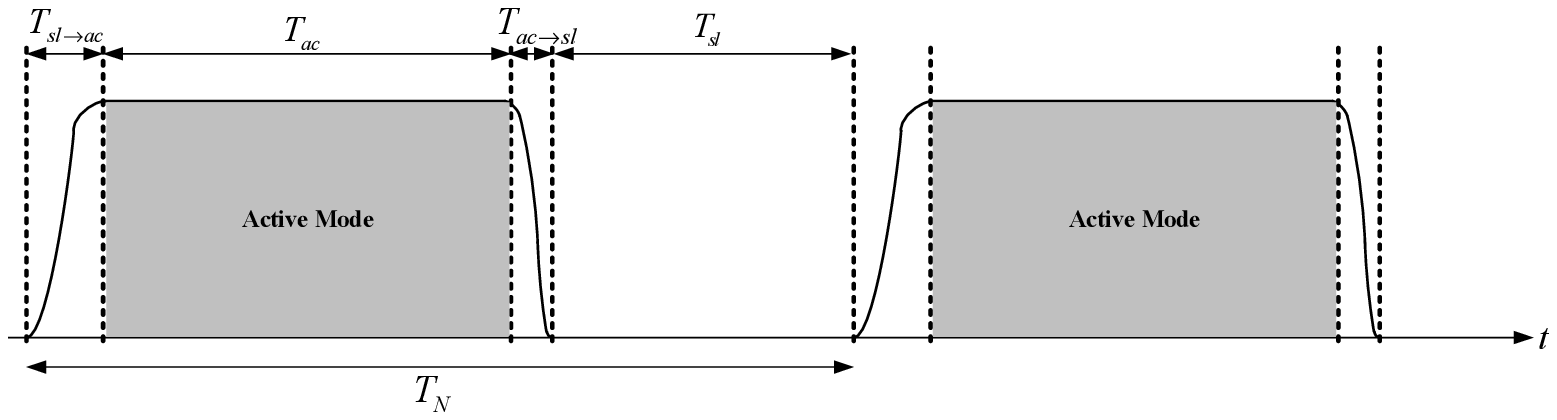,width=5.25in}} \caption{A
practical multi-mode operation in a proactive WSN.} \label{fig:
Time-Basis}
\end{figure}

\begin{figure}[t]
\centerline{\psfig{figure=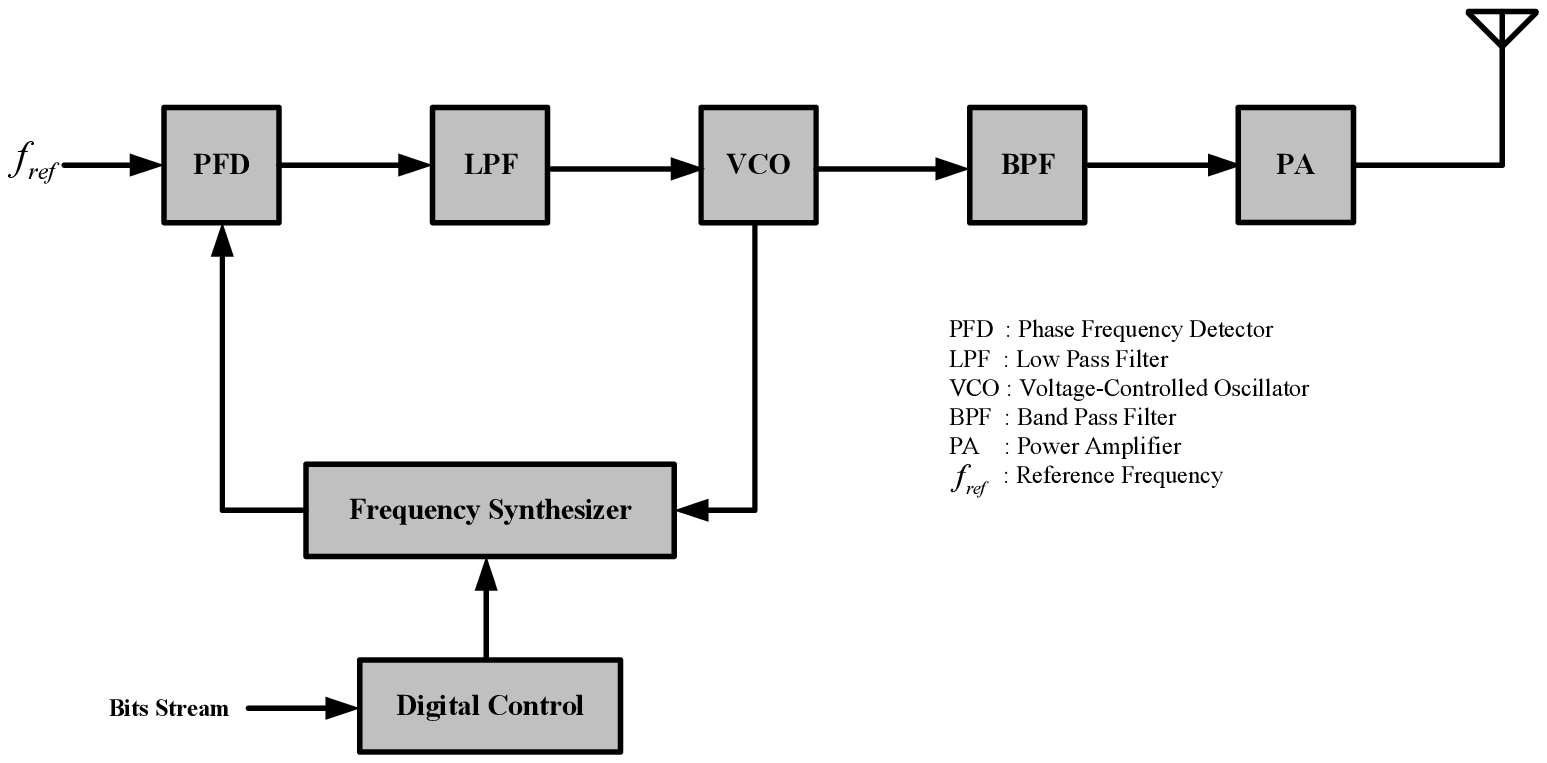,width=5.25in}}
\caption{General block diagram of an MFSK modulator.} \label{fig:
FSKmodulator}
\end{figure}

\begin{figure}[t]
\centerline{\psfig{figure=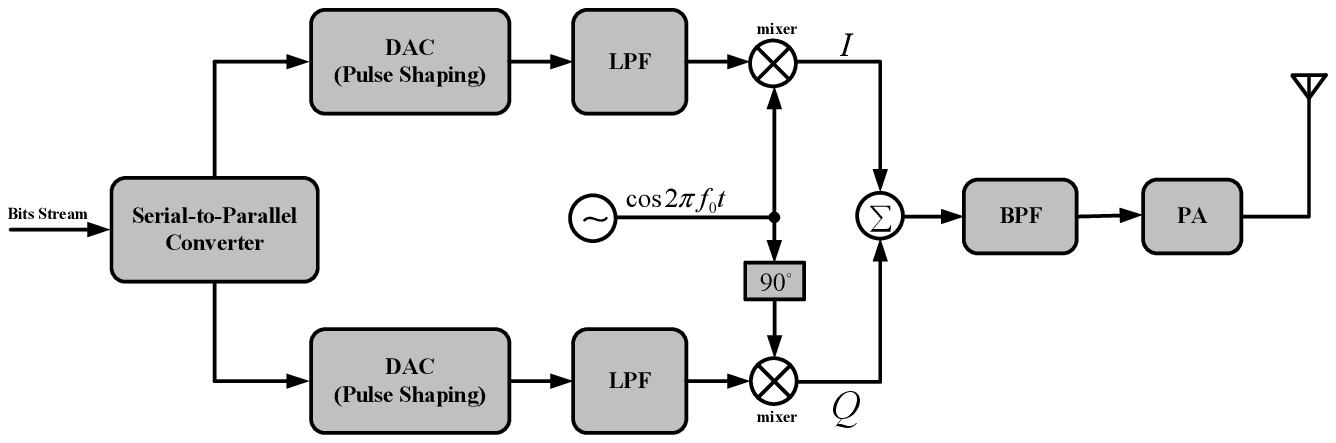,width=5.9in}}
\caption{General block diagram of an MQAM modulator.} \label{fig:
MQAM_Mod}
\end{figure}

\begin{figure}[t]
\centerline{\psfig{figure=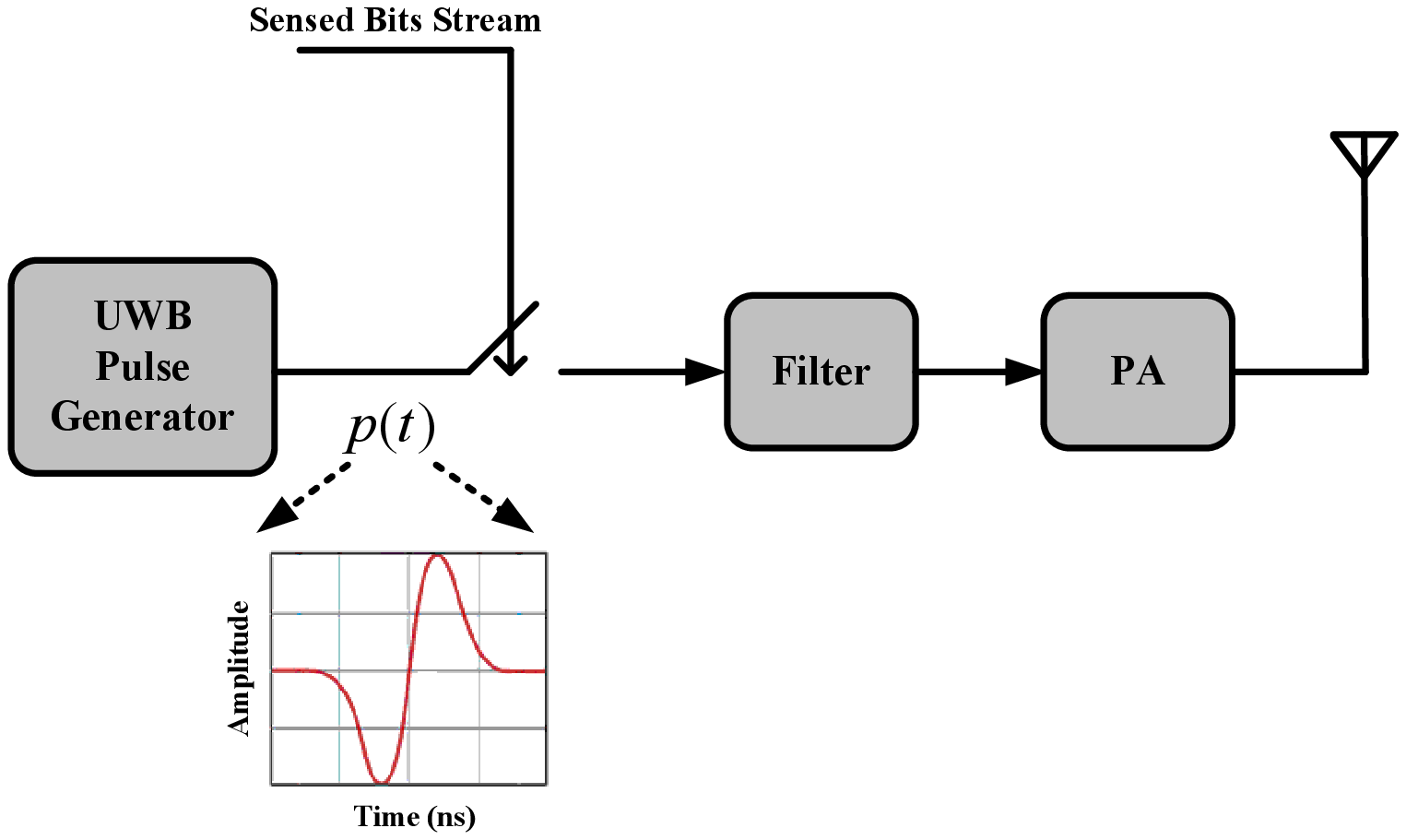,width=4.0in}}
\caption{Block diagram of an OOK transmitter  with Gaussian
monocycle $p(t)$ with duration $T_p$.} \label{fig: OOK_TX}
\end{figure}

\begin{figure}[t]
\centerline{\psfig{figure=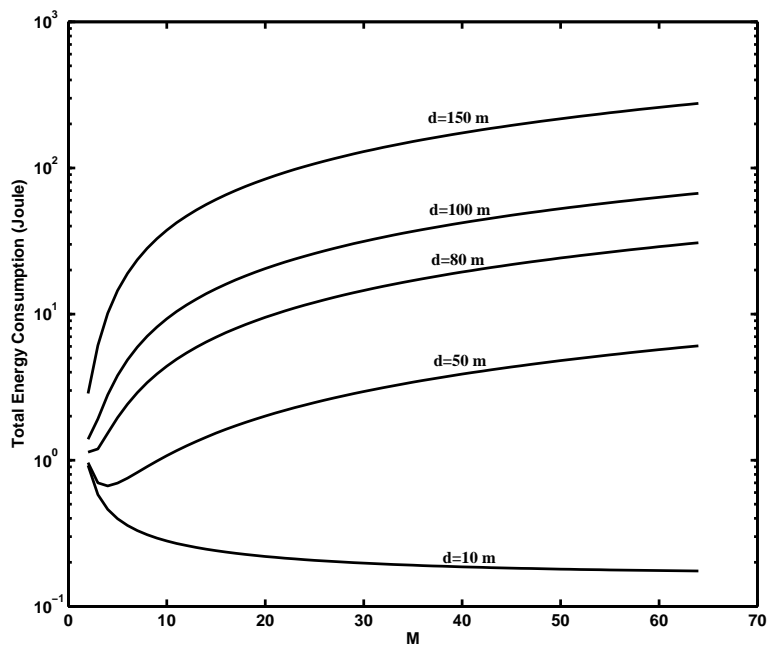,width=4.65in}}
\caption{Total energy consumption $\mathcal{E}_{N}^{QA}$ vs. $M$
over Rayleigh fading channel model with path-loss for
$P_s=10^{-3}$.} \label{fig: Total_Energy}
\end{figure}

\begin{figure}[bhpt]
\centerline{\psfig{figure=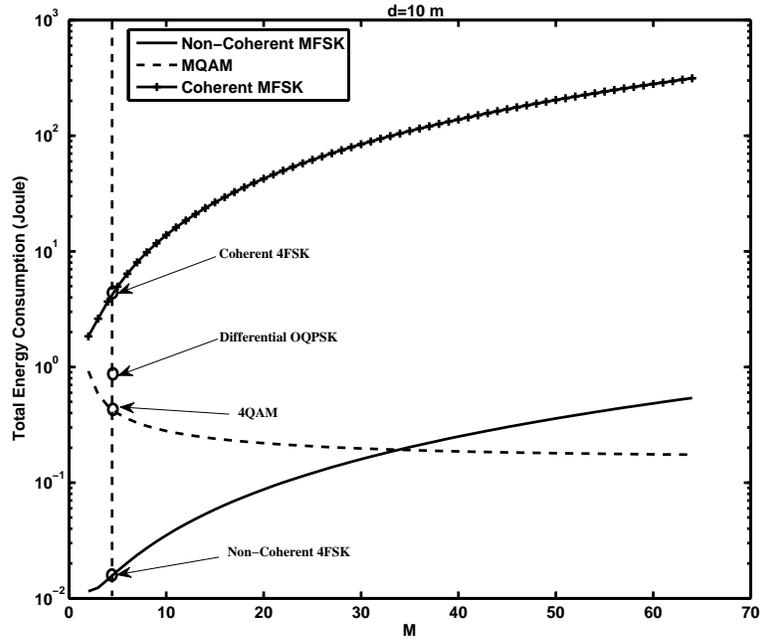,width=4.6in}}
\vspace{-7pt} \center{\hspace{16pt} \small{(a)}} \vspace{10pt}
\hspace{1pt}
\centerline{\psfig{figure=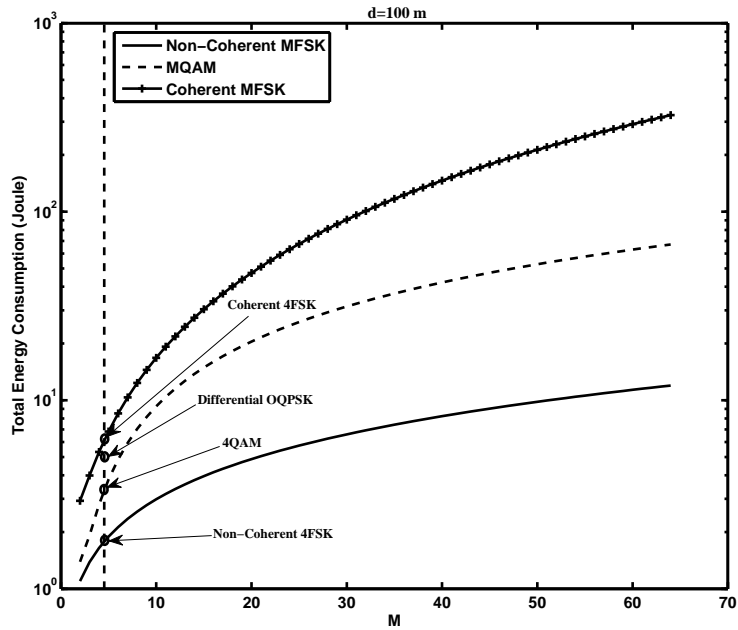,width=4.5in}}
\vspace{-35pt}
\center{\hspace{14pt} \small{(b)}} \\
\vspace{-7pt} \caption[a)  and b) .] { \small{Total Energy
consumption of transmitting $N$ bits vs. $M$ for MFSK, MQAM and
differential OQPSK over Rayleigh fading channel model with path-loss
and $P_s=10^{-3}$, a) $d=10$ m, and b) $d=100$ m.}} \label{fig:
Total_Energy_dm}
\end{figure}

\begin{figure}[t]
\centerline{\psfig{figure=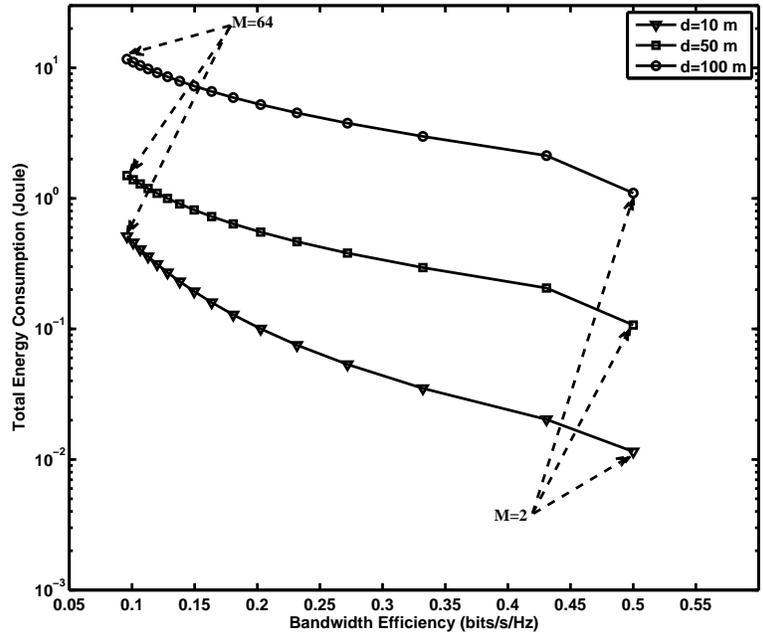,width=4.65in}}
\caption{Total energy consumption of transmitting $N$ bits versus
 bandwidth efficiency for NC-MFSK, and for $P_s=10^{-3}$ and different values of $d$ and $M$.} \label{fig: Energy_vs_Bandwidth}
\end{figure}

\begin{figure}[t]
\centerline{\psfig{figure=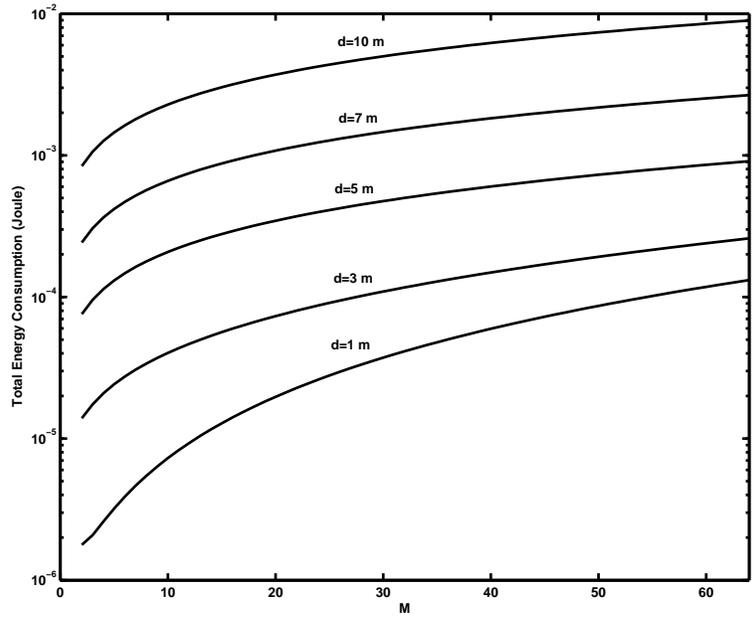,width=4.65in}}
\caption{Total energy consumption of transmitting $N$ bits versus
 $M$ for M-PPM over Rayleigh fading
channel model with path-loss and $P_s=10^{-3}$.} \label{fig:
Total_EnergyPPM_vs_M}
\end{figure}

\begin{figure}[t]
\centerline{\psfig{figure=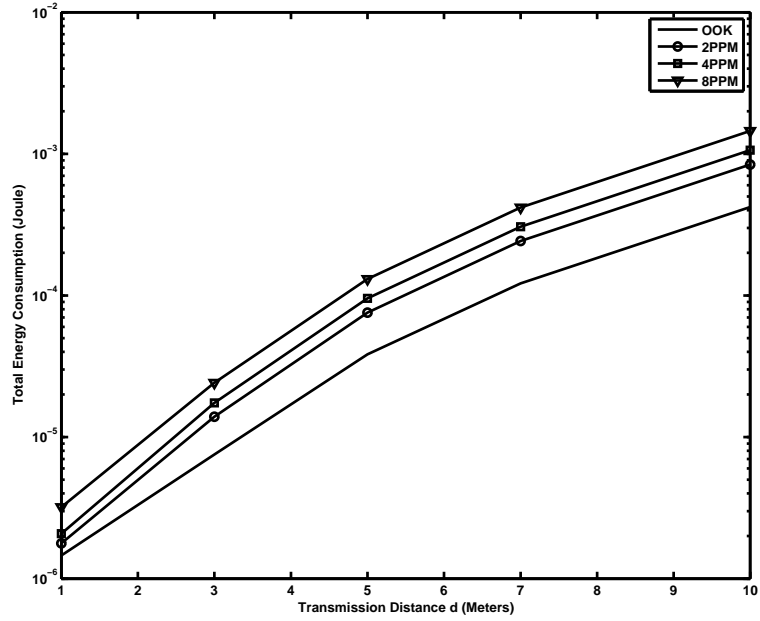,width=4.65in}}
\caption{Total energy consumption of transmitting $N$ bits versus
 $d$ for OOK and M-PPM with M=2,4,8, and for $P_s=10^{-3}$.} \label{fig: Total_Energy_UWB_vs_d}
\end{figure}

\begin{table}
\label{table01} \caption{System Evaluation Parameters
\cite{Karl_Book2005, Cui_GoldsmithITWC0905}} \centering
  \begin{tabular}{|lll|lll|}
  \hline
                         & Pass-Band                   &                          &              &    UWB             &                               \\
  \hline
   $N=8192$ bits         & $T_N=1.4$ sec               & $\mathcal{P}_{DAC}=7$ mw &  $N=20000$   &   $N_0=-180$ dB    &  $\mathcal{P}_{PG}=675$ $\mu$w \\

   $B=62.5$ KHz          & $N_0=-180$ dB               & $\mathcal{P}_{ADC}=7$ mw &  $B=500$ MHz &   $T_N=100$ msec   &  $\mathcal{P}_{LNA}=3.1$ mw    \\

   $M_l=40$ dB           & $\mathcal{P}_{Sy}=10$ mw    & $\mathcal{P}_{Mix}=7$ mw &  $M_l=40$ dB &   $T_{tr}=2$  nsec &  $\mathcal{P}_{ED}=3$ mw       \\

   $\mathcal{L}_1=30$ dB & $\mathcal{P}_{Filt}=2.5$ mw & $\mathcal{P}_{ED}=3$ mw  & $\mathcal{L}_1=30$ dB & $\mathcal{P}_{Filt}=2.5$ mw  & $\mathcal{P}_{ADC}=7$ mw \\

   $\eta=3.5$            & $\mathcal{P}_{Filr}=2.5$ mw & $\mathcal{P}_{IFA}=3$ mw & $\eta=3.5$   &   $\mathcal{P}_{Filr}=2.5$ mw   & $\mathcal{P}_{Int}=3$ mw  \\

   $\Omega=1$            & $\mathcal{P}_{LNA}=9$ mw    &      &     &    &   \\

  \hline
  \end{tabular}
\end{table}

\begin{table}
\label{table022} \caption{Total Energy Consumption (in Joule) of
NC-MFSK, MQAM and OQPSK over Rician Fading Channel with Path-Loss}
\centering
  \begin{tabular}{|c|cccc|ccc|ccc|}
   \hline

   &          &        & $K=1$ dB &       &            & $K=10$ dB &       &        & $K=15$ dB &       \\
   \hline
   &  $M$     & OQPSK  & NC-MFSK  & MQAM  &   OQPSK    & NC-MFSK   & MQAM  & OQPSK  & NC-MFSK   & MQAM  \\
   \hline
   & 4        & 1.1241 & 0.0173   & 0.5621&    1.1241  & 0.0171    &0.5620 &  1.1241& 0.0171    & 0.5620\\

   d=10 m & 16         &        & 0.0769   & 0.2819&            & 0.0765    &0.2810 &        & 0.0765    & 0.2810\\

   & 64       &        & 0.6558   & 0.1924&            & 0.6545    &0.1874 &        & 0.6545    & 0.1874\\
   \hline
   \hline
   & 4          & 1.2236 & 0.5835   & 0.8873 &   1.1445   & 0.0194    &0.5652 & 1.1310 & 0.0175    & 0.5627\\

   d=100 m & 16         &        & 1.4920   & 3.2049 &            & 0.0785    &0.2989 &        & 0.0767    & 0.2843\\

   & 64         &        & 4.6199   & 16.1010&            & 0.6570    &0.2615 &        & 0.6547    & 0.2002\\
   \hline
  \end{tabular}
\end{table}

\end{document}